\documentclass[11pt]{article} 
\usepackage{amsfonts}
\usepackage[centertags]{amsmath} 
\usepackage{amscd, euscript, bbm}
\newcommand{\bb}{\begin {minipage} {3cm}\begin{center}}
\newcommand{\ee}{\end{center}\end{minipage}}
\newcommand{\bc}{\begin {minipage} {2.5cm}\begin{center}}
\newcommand{\bd}{\end{center}\end{minipage}}

\newcommand{\R}{{\mathbb R}}
\newcommand{\Z}{{\mathbb Z}}

\newcommand{\G}{\mathfrak{G}}
\newcommand{\HH}{\mathfrak{H}}
\newcommand{\g}{\mathfrak{g}}
\newcommand{\h}{\mathfrak{h}}

\newcommand{\M}{{\cal M}}
\newcommand {\f}{\footnote}
\newcommand {\q}{\begin{quote} \small}
\newcommand {\be}{\begin {eqnarray}}
\newcommand {\e}{\end {eqnarray}}

\newcommand{\rmod}{\,{\textup{mod}}}
\newcommand{\Ord}{{\cal{O}}}
\newcommand{\hgs}{{\hat{\g}(\s)}}
\newcommand{\hhs}{{\hat{\h}(\s)}}
\def\hj{\hat{J}}
\newcommand{\idop}{\ensuremath{\mathbbm{1}}}
\newcommand{\hsig}{h_{\sigma}}

\def\mnnrnsrs{{m+n+\srac{n(r)+n(s)}{\r(\s)}}}
\def\mnnrnsrsf{{m+n+\frac{n(r)+n(s)}{\r(\s)}}}
\def\mnrrs{{m+\srac{n(r)}{\r(\s)}}}

\def\0b{\ }

\def\Nrm{{N(r)\m}}

\def\Nsn{{N(s)\n}}
\def\Ntd{{N(t)\d}}

\def\a{\alpha}

\def\bea{\begin{eqnarray}}
\def\beas{\begin{eqnarray*}}

\def\d{\delta}
\def\dual{\underset{\s}{\longrightarrow}}

\def\ee{\end{equation}}
\def\eea{\end{eqnarray}}
\def\eeas{\end{eqnarray*}}

\def\gfrak{\mbox{$\mathfrak g$}}
\def\hfrak{\mbox{$\mathfrak h$}}

\def\ginv{g^{-1}}

\def\hc{^\dagger}
\def\hcj{\dagger}

\def\hg{\hat{g}}

\def\hj{\hat{J}}
\def\hjb{{\hat{\bar{J}\hspace{.05in}}\hspace{-.05in}}}

\def\hjs{{\hat{j\hspace{.03in}}\hspace{-.03in}}}

\def\l{\lambda}
\def\m{\mu}
\def\mnnrnsrs{{m+n+\srac{n(r)+n(s)}{\r(\s)}}}
\def\mnnrnsrsf{{m+n+\frac{n(r)+n(s)}{\r(\s)}}}
\def\mnrn{{-n(r),\n}}
\def\mnrrs{{m+\srac{n(r)}{\r(\s)}}}

\def\n{\nu}
\def\nn{\nonumber}
\def\nnsrs{n+\srac{n(s)}{\r(\s)}}

\def\nrm{{n(r)\m}}

\def\nsn{{n(s)\n}}

\def\pl{\partial}

\def\r{\rho}
\def\s{\sigma}

\def\sG{{\cal G}}

\def\scf{{\cal F}}

\def\sg{\smal{\EuScript{G}}}

\def\shh{\smal{\EuScript{H}}}

\def\sm{{\cal M}}

\def\srac#1#2{\smal{\frac{#1}{#2}}}
\def\srange{\s=0,\ldots,N_c-1}
\def\st{{\cal T}}

 \def\smal#1{\mbox{\small $#1$}}
 \def\big#1{\mbox{\large $#1$}}
 \def\Big#1{\mbox{\Large $#1$}}

\numberwithin{equation}{section}

\begin{document}
\newtheorem {lemma}{Lemma}[subsection]
\newtheorem {theorem}{Theorem}[subsection]
\newtheorem {coro}{Corollary}[subsection]
\newtheorem {defi}{Definition}[subsection]
\newtheorem {obs}{Remark}[subsection]
\newtheorem {prop}{Proposition}[subsection]
\newtheorem {exa} {Example} [subsection]

\begin{titlepage}
\begin{center}

\today           \hfill UCB-PTH-02/17   \\
                               \hfill LBNL-50134    \\
\vskip .25in
\def\thefootnote{\fnsymbol{footnote}}
{\Large \bf The General Coset Orbifold Action}
\vskip 0.5in

M. B. Halpern\footnote{halpern@physics.berkeley.edu} and F. Wagner\footnote{fwagner@lbl.gov}

\vskip 0.2in
{\em Department of Physics, University of California, Berkeley, CA 94720-7300, USA\\
and \\
     Theoretical Physics Division, Lawrence Berkeley National Laboratory\\
     Berkeley, CA 94720, USA}
        
\end{center}

\vskip .3in

\vfill

\begin{abstract}
Recently an action formulation, called the general WZW orbifold action, was given for each sector of every WZW orbifold. In this paper we gauge this action by general twisted gauge groups to find the action formulation of each sector of every coset orbifold. Connection with the known current-algebraic formulation of coset orbifolds is discussed as needed, and some large examples are worked out in further detail. 
\end{abstract}

\vfill

\end{titlepage}

\setcounter{footnote}{0}


\pagebreak
\section{Introduction}\label{sect:intro}
As string theory slows for a third time, a quiet revolution is taking place in the local theory of {\em current-algebraic orbifolds}. In particular, Refs.\,[1--3] 
gave the twisted currents and stress tensor in each twisted sector of any current-algebraic orbifold $A(H)/H$, where $A(H)$ is any current-algebraic conformal field theory [4--10]
with a finite symmetry group $H$.

More recently, the special case of the {\em WZW orbifolds} $A_g(H)/H$ was worked out in further detail \cite{deBoer:MBH01,MBHObers}, while extending the operator algebra in this case to include the {\em twisted affine primary fields}, {\em twisted vertex operator equations} and {\em twisted Knizhnik-Zamolodchikov equations} for each sector of every WZW orbifold:
\begin{itemize}
\item The WZW permutation orbifolds \cite{deBoer:MBH01,MBHObers}
\item The inner-automorphic WZW orbifolds \cite{deBoer:MBH01}
\item The (outer-automorphic) charge conjugation orbifold on $\mathfrak{su}(n)$   \cite{MBHObers}
\item Other outer-automorphic WZW orbifolds on simple g \cite{MBHObers}
\end{itemize} 
Ref.\,\cite{MBHObers} also solved the twisted vertex operator equations in an abelian limit to obtain the twisted vertex operators for each sector of a large class of abelian orbifolds.

In addition to the operator formulation, a {\em general WZW orbifold action} was also given in Ref.\,\cite{deBoer:MBH01}, with applications to special cases in Refs.\,\cite{deBoer:MBH01,MBHObers}. The general WZW orbifold action provides the classical description of each sector of every WZW orbifold in terms of appropriate {\em group orbifold elements}, which are the classical limit of the twisted affine primary fields.

In this paper, we gauge the general WZW orbifold action by general twisted gauge groups to obtain the {\em general coset orbifold action}, which is the classical description of each sector of every coset orbifold $A_{g/h}(H)/H$. 

The highlights of the paper are as follows. With the help of a new {\em twisted Polyakov-Wiegmann identity} for each twisted subgroup in each sector of any WZW orbifold, the general coset orbifold action is obtained in Sec.\,\ref{sect:cosetaction}.  In Sec.\,\ref{sect:twistedsubal} we give a {\em recipe} for going from the known operator formulation in each sector of any particular coset orbifold \cite{Evslin:1999ve, Halpern:2000vj} to the corresponding set of coset orbifold actions for that orbifold. The recipe is illustrated in Secs.\,\ref{sect:permcoset}-\ref{sect:other_ex}, where we work out a number of large examples in further detail --- including in particular the {\em general coset permutation orbifolds} in Sec.\,\ref{sect:permcoset}.

The paper is modestly self-contained, with background material introduced as needed. In particular, the general WZW orbifold action and its form for the WZW permutation orbifolds are reviewed in Secs.\,\ref{sect:action} and \ref{sectperm} respectively, and the operator formulation of the general coset orbifold is reviewed in Secs.\,\ref{sect:cosetaction} and \ref{sect:twistedsubal}.

\section{Background: The General WZW Orbifold Action}\label{sect:action}

Recently {\em the  general WZW orbifold action}, in terms of so-called {\em group orbifold elements} with definite monodromy, was given in Ref.\,\cite{deBoer:MBH01}. This action gives the classical formulation of all sectors of each WZW orbifold
\begin{gather}
\frac{A_g(H)}{H} \; , \; H \subset Aut(g) \,,  \qquad   g = \bigoplus_{I=0}^{K-1} \g^I\,
\end{gather}
where $A_g(H)$ is any WZW model on compact semisimple Lie $g$ with a finite symmetry group $H$. In this section we review some properties of this action which will be needed in our study of coset orbifolds below. 

The explicit form of the general WZW orbifold action \cite{deBoer:MBH01} on the cylinder $(t \in \R, 0 \leq \xi < 2\pi)$ and the solid cylinder $\Gamma$ is ($\partial_\pm = \partial_t \pm \partial_\xi$)
\begin{align}
 & \widehat{S}_{\hg(\s)} [{\cal{M}},\hg ] \equiv   \widehat{S}_{\hg(\s)} [{\cal{M}}(\st(T,\s)), \hg(\st(T,\s),\s) ]\nn
\end{align}
\begin{gather}
\!\! = -\frac{1}{8\pi}\int d^2\xi
 \0b \widehat{\rm Tr}\big{\{}\;\sm(\st(T,\s),\s)\0b\hat{g}^{-1} (\st(T,\s),\s)
\pl_+\hat{g}(\st(T,\s),\s)\0b\hat{g}^{-1}
(\st(T,\s),\s)\pl_-\hat{g}(\st(T,\s),\s)\;\big{\}}\nn\\
\;-\frac{1}{12\pi}\int_{\Gamma}\!\! \widehat{\rm Tr}\big{\{}\sm(\st(T\!,\s),\s)\0b(\hat{g}^{-1}
  (\st(T\!,\s),\s) d\hat{g}(\st(T\!,\s),\s)\;)^3\,\big{\}}, \quad \forall\,T\quad \text{for each } \s = 0, \ldots , N_c -1\, \label{gorbact}
\end{gather}
\begin{equation}
    (\hat{g}^{-1} d\hat{g})^3 = dt\,d\xi\,d\r\;\epsilon^{ABC}(\hat{g}^{-1} \pl_A\hat{g})(\hat{g}^{-1} \pl_B\hat{g})(\hat{g}^{-1} \pl_C\hat{g}),\quad
        \{A,B,C\}=\{t,\xi,\r\} \, \quad  d^2\xi\equiv dt\,d\xi \, .\nn
\end{equation}
Here $\s$ is the sector number of the WZW orbifold $A_g(H)/H$, $N_c$ is the number of conjugacy classes of $H$, $\st \equiv  \st(T,\s)$ are the {\em twisted representation matrices} which are functions of the untwisted representation matrices $T$ of Lie $g$, and ${\cal M}\equiv {\cal M}(\st,\s)$ is the invertible {\em twisted data matrix}. For each sector $\s$ and all $T$ of each WZW orbifold $A_g(H)/H$, the general forms of the twisted representation matrices $\st(T,\s)$ and the twisted data matrices $\M(\st,\s)$ are given in Ref.\,\cite{deBoer:MBH01}. As examples, the explicit forms of these twisted matrices are reviewed for the WZW permutation orbifolds in Sec.\,\ref{sectperm}. More generally, we will return to the twisted representation matrices in  Sec.\,\ref{sect:twistedsubal}. 

The quantities $\hat{g} \equiv \hat{g}({\cal T}(T,\s), \xi, t, \s)$ in (\ref{gorbact}) are the  group orbifold elements in twisted representation $\st$ of sector $\s$. The group orbifold elements are unitary matrices which form an infinite-dimensional Lie group $\widehat{\G}(\s)$ at each time $t$ in each sector $\s$ of the orbifold
\be
\widehat{\G}(\s) \equiv \times_\xi \G_\xi(\s), \quad \G_\xi(\s) \cong \G(\s),  \quad\s = 0, \dots, N_c-1
\e
where $\widehat{\G}(\s)$ is a formal product over  finite-dimensional Lie groups $\G_\xi(\s)$ at each spatial point $\xi$. The WZW orbifold action $\widehat{S}_{\hat{g}(\s)}$ is labelled by the infinite-dimensional Lie algebra
\be
\hat{g}(\s) = \bigoplus_\xi g_\xi(\s), \quad g_\xi(\s) \cong g(\s) \, ,  \quad\s = 0, \dots, N_c-1
\e
which generates $\widehat{\G}(\s)$.

In fact $\widehat{\G}(\s)$ is a {\em twisted infinite-dimensional Lie group} because the group orbifold elements exhibit the monodromy \cite{deBoer:MBH01}
\begin{subequations}\label{gmon}
\begin{align}
\hat{g}(\st, \xi + 2\pi, t, \s) & = E(T, \s) \hat{g}(\st, \xi, t, \s) E(T, \s)^*\, \label{gmonE}\\
E(T, \s)_{N(r) \mu}{}^{N(s)\nu}  & = \delta_\mu{}^\nu \delta_{N(r)- N(s),0\rmod R(\s)} e^{-2\pi i \frac{N(r)}{R(\s)}}\label{Glmon}\\
\hg (\st,\xi + 2\pi,t,\s)_{\Nrm}{}^{\Nsn}  &=   e^{-2 \pi i
\frac{N(r)-N(s)}{R(\s)}} \hg (\st,\xi,t,\s)_{\Nrm}{}^{\Nsn}\, 
\end{align} 
\end{subequations}
upon circumnavigation of the cylinder.
In Eq.(\ref{gmon}) the quantity  $R(\s) \equiv R(T,\s)$ is the order of the automorphism group element $h_\s \in H$ in untwisted matrix representation $T$ of Lie $g$, while the set of {\em spectral integers} $N(r) \equiv N(r,T,\s)$ and the {\em degeneracy indices} $\mu\equiv \mu(r,T,\s)$ are computed from the {\em extended $H$-eigenvalue problem} \cite{deBoer:MBH01} for untwisted representation $T$. The operation $\widehat{Tr}$ in (\ref{gorbact}) is the trace in the space of the unitary matrices
\be 
 \widehat{Tr}\{AB\} \equiv \sum_{r,\m,s,\n} A_\Nrm{}^\Nsn B_\Nsn{}^\Nrm \;.  
\e
In this formulation, all quantities are periodic $N(r) \rightarrow N(r) \pm R(\s)$ in their spectral integer labels and the spectral  integers can be pulled back into a fundamental range $\bar{N}(r) \in \{ 0, \dots, R(\s)-1\}$. These integers and indices are reviewed for the WZW permutation orbifolds \cite{deBoer:MBH01}  in Sec.\,\ref{sectperm}, and this data is given for the inner-automorphic WZW orbifolds and a class of outer-automorphic WZW orbifolds in Refs.\,\cite{deBoer:MBH01, MBHObers}.

 The two-sided monodromy (\ref{gmon}) of the group orbifold elements guarantees that the product of any two group orbifold elements, which is another group orbifold element, has the same monodromy. Moreover, the group orbifold elements are the {\em classical limit of the twisted affine primary fields} \cite{deBoer:MBH01} of the WZW orbifolds, and the associated twisted infinite-dimensional Lie algebra $\hat{g}(\s)$ is the action analogue of the underlying {\em twisted affine Lie algebra}  of the WZW orbifold. The tangent space description of the group orbifold elements is discussed in Sec.\,\ref{sect:twistedsubal}.

To see that the general WZW orbifold action (\ref{gorbact}) has trivial monodromy \cite{deBoer:MBH01}
\begin{gather}
\widehat{S}_{\hg(\s)}[{\cal M}(\st,\s), \hat{g}(\st, \xi + 2\pi, t,\s)] = \widehat{S}_{\hg(\s)}[{\cal M}(\st, \s), \hat{g}(\st, \xi, t,\s)] \, ,  \quad\s = 0, \dots, N_c-1
\end{gather}
 one needs the fact that $\hat{g}^{-1}$ has the same monodromy as that given for the group orbifold element $\hat{g}$ in Eq. (\ref{gmon}),  and the \emph{selection rule}\,\cite{ deBoer:1999na,  Halpern:2000vj, deBoer:MBH01} of the twisted data matrix \cite{deBoer:MBH01}
\begin{gather}
\label{eq:msel}
 E(T,\s)\sm(\st,\s)E(T,\s)^\ast = \sm(\st,\s)\quad \Leftrightarrow \quad
 \sm(\st,\s)_{N(r)\m}{}^{N(s)\n}
 \left(1-e^{-\frac{2\pi i (N(r)-N(s))}{R(\s)}}\right) = 0 . 
\end{gather}
The ${\cal M}$-selection rule (\ref{eq:msel}) is a reflection in the WZW orbifold of the $H$-symmetry of $A_g(H)$. 

The invertibility of ${\cal M}(\st, \s)$ and the vanishing commutator \cite{deBoer:MBH01} 
\begin{equation}
 [\sm(\st,\s),\hat{g}(\st,\xi,t,\s)] =0 \label{[sm,sg]=0}\,,  \qquad \forall \; \hat{g}(\st, \xi, t, \s) \in \widehat{\G}(\s) 
\end{equation}
are central in obtaining the equations of motion \cite{deBoer:MBH01}  of sector $\s$ of $A_g(H)/H$:
\begin{subequations}\label{EOMtwis}
\begin{gather}
\pl_-\hj(\st,\xi,t,\s) = \pl_+\hjb(\st,\xi,t,\s) = 0
  \label{twistedEOMJmatrix} \\
\hj(\st\!,\xi,t,\s) \equiv\! -\frac{i}{2}\hat{g}^{-1}(\st\!,\xi,t,\s)\pl_+
  \hat{g}(\st\!,\xi,t,\s)\,, \quad
 \hjb(\st\!,\xi,t,\s) \equiv\! -\frac{i}{2}\hat{g}(\st\!,\xi,t,\s)\pl_-\hg^{-1}
 (\st\!,\xi,t,\s) \label{tw_g_currents}\\
\hj(\st\!, \xi+2\pi, t, \s) = E(T, \s) \hj(\st\!, \xi, t, \s) E(T,\s)^*  , \;
\hjb(\st\!, \xi+2\pi, t, \s) = E(T, \s) \hjb(\st\!, \xi, t, \s) E(T,\s)^*  .\label{tw_curr_mono}
\end{gather}
\end{subequations}
Here $\hj$ and $\hjb$ are the twisted classical left- and right-mover matrix currents, whose monodromies (\ref{tw_curr_mono}) follow from the monodromy of the group orbifold elements in Eq. (\ref{gmon}).

In the untwisted sector $\s = 0$ of each WZW  orbifold, the general WZW orbifold action (\ref{gorbact}) reduces to the ordinary, untwisted WZW action for $A_g(H)$ on compact semisimple $g$ with a symmetry group $H$:
\begin{subequations}\label{untw_WZW}
\begin{gather}
\G = \G(0)\cong \G_\xi(0)\,,\quad  g= g(0) =  \bigoplus_{I=0}^{K-1} \g^I\, ,  \quad S_{WZW} \equiv S_{\hg(0)}\,\label{11a} \\
T \equiv \st(T, 0), \quad   M(k,T) \equiv  \sm (\st, 0)\,  \\
 g(T, \xi, t)\equiv   \hat{g}(\st, \xi, t, 0) \, , \quad [M(k,T), g(T, \xi, t)] = 0\label{11c}
\end{gather}
\begin{gather}
T= \bigoplus_{I=0}^{K-1} T^I, \quad 
g(T, \xi, t) = \bigoplus_{I=0}^{K-1} g_I(T^I, \xi, t)\, , \quad  M(k,T) = \bigoplus_{I=0}^{K-1} \Bigg(\frac{k_I}{y_I(T^I)} \idop_I\Bigg)\,\label{11d} 
\end{gather}
\begin{equation}
\!\!\!\!\!\!\!\!\!\!\!\!\!\!\!\! S_{WZW}[M,g] =-\frac{1}{8\pi}\int d^2\xi\0b {\rm Tr}\Big{(}M(k,T)\0b\ginv (T)\pl_+
 g(T)\0b\ginv(T)\pl_-g(T)\Big{)}
 \end{equation}
 $$
    \quad\quad\quad\quad\quad-\frac{1}{12\pi}\int_{\Gamma} {\rm Tr}\Big{(}M(k,T)\0b
    (\ginv(T)dg(T))^3\0b\Big{)} \quad \label{action-a}
$$
\begin{equation}
\quad\quad\quad\quad\quad\quad\quad\quad =-\frac{1}{8\pi}\sum_I \frac{k_I}{y_I(T^I)}\int d^2\xi \ {\rm Tr}\Big{(}
g_{I}^{-1} (T^I)\pl_+g_I(T^I)\0b g_{I}^{-1}(T^I)\pl_-g_I(T^I)\Big{)} +  \ldots \, .
\end{equation}
\end{subequations}
Here $g(T, \xi, t)$ is an element of the Lie group $\G_\xi(0) \cong \G$ in untwisted representation $T$ of the untwisted Lie algebra $g$, and $M(k,T)$ is the untwisted data matrix\f{As an example, permutation invariant systems satisfy $\g^I \cong \g$, $k_I = k$, $T^I =  T$ and $y_I = y(T)$ so that $M(k,T) = (k/y(T)) \idop$ in this case.}  --- which involves the levels of $\{\g^I\}$ and the Dynkin indices $y_I$ of the representations $T^I$ of $\{\g^I\}$. The vanishing commutator $[M, g]=0$ in (\ref{11c}) is a consequence of the block structure shown in (\ref{11d}). The action label $\hat{g}(0)$ in (\ref{11a}) is the action analogue of the underlying untwisted affine Lie algebra of the untwisted WZW model.

In the untwisted action formulation, the $H$-symmetry of $A_g(H)$ is encoded in the $H$-symmetry of the untwisted data matrix \cite{deBoer:MBH01} 
\begin{gather}
 W^\hcj(h_\s;T)M(k,T)W(h_\s;T) = M(k,T),\quad \forall \ h_\s \in H
  \subset  Aut(g)  \label{H_sym}
\end{gather}
where $W(\hsig ; T)$ is the action of $\hsig \in H$  in representation $T$. The ${\cal M}$-selection rule (\ref{eq:msel}) is the orbifold dual of the $H$-symmetry (\ref{H_sym}) and the extended $H$-eigenvalue problem \cite{deBoer:MBH01} 
\begin{subequations}\label{exH}
\begin{gather}
W(\hsig ; T) U^{\dagger}(T, \s) = U^{\dagger}(T, \s) E(T, \s), \quad W(\hsig; T) \in H, \quad \s = 0, \dots, N_c-1\\
W(\hsig; T) ^{R(T,\s)} = 1
\end{gather}
\end{subequations}
is defined for $W(\hsig ; T)$.
The order $R(T,\s)$ of $W(\hsig; T)$ and the eigenvalue matrix $E(T, \s)$ have already appeared in Eqs. (\ref{gmon}), (\ref{eq:msel}) and (\ref{EOMtwis}). Then it is easily checked that the untwisted WZW action (\ref{untw_WZW}) is invariant under the symmetry group
\begin{subequations}\label{155}
\begin{gather}
S_{\rm WZW}[ M,g'] = S_{\rm WZW}[ M,g]\\
g(T,\xi,t)' = W(h_\s;T) g(T,\xi,t) W\hc (h_\s;T) \in \G_\xi(0), \quad \G_\xi(0) \cong \G\label{13b}
\end{gather}
\end{subequations}
because the untwisted data matrix is $H$-invariant. In a slight abuse of notation, we may write untwisted relations such as (\ref{13b}) in the form
\be\label{215a}
g(T, \xi, t)' \in \G
\e
which parallels the conventional description of  untwisted Lie groups and Lie algebras in WZW models.

The general WZW orbifold action (\ref{gorbact}) was derived in Ref.\,\cite{deBoer:MBH01} from the untwisted $H$-symmetric WZW system by the method of {\em eigenfields} and {\em local isomorphisms} $(\dual)$
\begin{subequations}
\begin{gather}
{{\EuScript G}}(T,\xi, t) \equiv U(T, \s) {g} (T, \xi, t) U^\dagger(T, \s)\, , \\
{{\EuScript G}}(T,\xi, t)' = E(T, \s) {\EuScript G} (T, \xi, t) E(T, \s)^*\\
{\cal M}(\st, \s) \equiv U(T, \s) M(k,T) U^\dagger(T, \s) \\
{\EuScript G} \dual \hat{g}, \\
\text{automorphic response } E(T, \s) \dual \text{monodromy } E(T, \s) \\
\hat{g}(\st, \xi+2\pi, t, \s) = E(T, \s) \hat{g}(\st, \xi, t, \s) E(T, \s)^*\,    
\end{gather}
\end{subequations}
which uses the eigendata of the extended $H$-eigenvalue problem in (\ref{exH}). We mention in particular that the twisted result $[{\cal M}, \hat{g}] = 0$ in (\ref{[sm,sg]=0}) follows by local isomorphisms from the untwisted relation $[M, g] = 0$ in (\ref{11c}).

\section{About The Stress Tensors of $A_g(H)/H$ and $A_{g/h}(H)/H$}\label{sectbt}

In this section we set our stage by reviewing in broad terms what is known [1--3, 11,12] 
about the twisted currents and stress tensors of the WZW orbifolds and the coset orbifolds. We begin with the twisted left-mover currents $\hj_{\hgs} = \{{\hat J}_{n(r) \mu} (\xi, t,\s)\}$ of the {\em general twisted current algebra} $\hgs$, which is the operator analogue of the twisted infinite-dimensional Lie algebra $\hat{g}(\s)$ at the action level. The mode form of the general twisted current algebra $\hgs$ is reviewed in  Sec.\,\ref{sect:twistedsubal}. 

In sector $\s$ of any WZW orbifold, the monodromies of the twisted currents $\hj_{\hgs}$ 
\begin{subequations}\label{current_components}
\begin{gather}
{\hat J}_{n(r) \mu} (\xi + 2 \pi, t,\s) = E_{n(r)}(\s)
{\hat J}_{n(r) \mu} (\xi, t,\s)\quad \leftrightarrow \quad
{\hat J}_{n(r) \mu} (ze^{2\pi i},\s) =E_{n(r)}(\s) 
{\hat J}_{n(r) \mu} (z,\s) \,\\
E_{n(r)}(\s) = e^{-2\pi i \frac{n(r)} {\rho(\s)}},  \quad\s = 0, \dots, N_c-1\label{monoex} 
\end{gather} 
\end{subequations}
are given in Refs.\,\cite{deBoer:1999na, Halpern:2000vj}.
Here the monodromies on the sphere are equivalent to those on the cylinder, and these monodromies are consistent (see Sec.\,\ref{sect:twistedsubal}) with the monodromy of the left-mover classical matrix current in (\ref{EOMtwis}). The quantity $\rho(\s)$ is the order of $h_\s \in H$ when $\hsig$ acts in the adjoint representation of Lie $g$, and the set of {\em spectral integers} $n(r) \equiv n(r,\s)$ and the {\em degeneracy indices} $\mu\equiv \mu(r, \s)$ are computed from the {\em $H$-eigenvalue problem} \cite{deBoer:1999na, Halpern:2000vj}, which is a special case of the extended $H$-eigenvalue problem. As seen in (\ref{monoex}), all quantities are also periodic $n(r) \rightarrow n(r) \pm \rho(\s)$ and these spectral integers can be pulled back into the fundamental range $\bar{n}(r) \in \{0, \dots, \rho(\s)-1\}$.

The corresponding left-mover stress tensor of sector $\s$ of $A_g(H)/H$ is the {\em twisted affine-Sugawara construction}
\begin{subequations}
\begin{gather}
\hat{T}_{\hgs} (z) = 
{\cal L}_{\hgs}^{\nrm ;  \mnrn} (\s)
: \hj_{\nrm} (z,\s) \hj_{\mnrn} (z,\s) :\,\\
  \hat{T}_{\hgs}(ze^{2\pi i }) = \hat{T}_{\hgs}(z) , \;\quad \hat{c}_{\hgs} = c_g,  \quad\s = 0, \dots, N_c-1\,
\end{gather}
\end{subequations}
where $: \cdot :$ is operator product normal ordering \cite{Evslin:1999qb,deBoer:1999na,  Halpern:2000vj} and  ${\cal L}_{\hgs}(\s)$ is called the {\em twisted inverse inertia tensor} of sector $\s$. For each sector $\s$ of each WZW orbifold $A_g(H)/H$, the general form of ${\cal L}_{\hgs}(\s)$ is given in Ref.\,\cite{deBoer:MBH01}. The corresponding twisted right-mover currents and twisted right-mover affine-Sugawara construction are discussed in Ref.\,\cite{deBoer:MBH01}. In the untwisted sector $\s = 0$, the twisted affine-Sugawara constructions reduce to the ordinary left- and right-mover affine-Sugawara constructions \cite{Bardakci:1971nb,Halpern:1971ay, kz, segalunpub} of the symmetric WZW model $A_g(H)$. As an example, the stress tensors of the WZW permutation orbifolds are reviewed in Sec.\,\ref{sectperm}.

Our task in this paper is to use the general orbifold WZW action (\ref{gorbact}) as a starting point to obtain the  action formulation of the {\em general coset orbifold} \cite{Evslin:1999ve,Halpern:2000vj}
\begin{gather}
\frac{A_{\frac{g}{h}}(H)}{H}\,,  \quad h \subset g\, ,  \quad H \subset Aut(g)\, , \quad H \subset Aut(h) \,. 
\end{gather}
Here $A_g(H)$ is an  $H$-invariant WZW model on $g$, $h \subset g$ is an {\em $H$-covariant subalgebra} of $g$ and the coset construction in this case
\be
\frac{g}{h} = \frac{g(H)}{h(H)} = \frac{g}{h}(H)
\e
is called an {\em $H$-invariant coset construction}. It is known \cite{Evslin:1999ve,Halpern:2000vj} that $g(H)$ and $h(H)\subset g(H)$ map under local isomorphisms to the twisted algebra $\hat{\g}(\s)$ and the twisted subalgebra $\hat{\mathfrak h}(\s)$ 
\be\label{lim}
g\dual \hat{\g}(\s), \quad h \dual \hhs \subset \hgs, \quad \frac{g}{h} \dual \frac{\hgs}{\hhs}
\e
and it is also known \cite{Evslin:1999ve,Halpern:2000vj} that the left-mover stress tensor of sector $\s$ of the general coset orbifold is a difference of twisted affine-Sugawara constructions
\begin{subequations}
\begin{gather}\label{stressT_gauge}
\hat{T}_{\srac{\hat{\mathfrak{g}}(\s)}{\hat{{\mathfrak{h}}}(\s)}} (z) = \hat{T}_{\hat{\mathfrak{g}}(\s)} (z) - \hat{T}_{\hat{\mathfrak{h}}(\s)} (z) \,,\quad  \hat{\mathfrak{h}}(\s) \subset \hat{\mathfrak{g}}(\s)\, ,\quad \s = 0, \dots, N_c-1\, \\
\hat{c}_{\frac{\hgs}{\hhs}} = c_{g/h} = c_{g} - c_{h}
\end{gather}
\end{subequations}
for the WZW orbifolds $A_g(H)/H$ and $A_h(H)/H$. The form (\ref{stressT_gauge}) of the coset orbifold stress tensor also follows \cite{Evslin:1999ve,Halpern:2000vj} by the method of eigenfields and local isomorphisms from the familiar form $T_{g/h} = T_g - T_h$ of the $g/h$ coset construction \cite{Bardakci:1971nb, Halpern:1971ay, Goddard:1985vk}.  The generators of $\hhs$ are the modes of the {\em twisted $h$ currents} $\hj_\hhs(z,\s)$
\be
\hat{T}_{\frac{\hgs}{\hhs}}(z) \hj_{\hhs}(w, \s) = \Ord\big( (z-w)^0\big)\,, \quad \s =0, \dots, N_c-1 
\e
 which are the twisted $(0,0)$ operators of the coset orbifold stress tensor of sector $\s$. 

In a continuation of this discussion, Sec.\,\ref{sect:twistedsubal} reviews  the systematic approach \cite {deBoer:MBH01,MBHObers} to the identification of the twisted affine subalgebras $\{\hat{\mathfrak h}(\s) \subset \hat{\g}(\s), \s = 0, \dots, N_c-1\}$ of each particular coset orbifold $A_{g/h}(H)/H$. Using these operator embeddings, we will also discuss in that section how to identify the action analogues $\{ \hat{h}(\s) \subset \hat{g}(\s), \s = 0, \dots, N_c-1\}$ of the twisted affine subalgebras.




\section{The General Coset Orbifold Action}\label{sect:cosetaction}
To obtain the action formulation of coset orbifolds, we will first learn to gauge the WZW orbifold action (\ref{gorbact}) by any particular twisted infinite-dimensional subgroup $\widehat{\HH}(\s) \subset \widehat{\G}(\s)$.

As a first step, we introduce the {\em subgroup orbifold elements} $\hat{h}(\st, \xi, t,\s)$ of any twisted infinite-dimensional subgroup $\widehat{\HH}(\s)$ 
\be
\hat{h}(\st, \xi, t, \s) \in \widehat{\HH}({\s}) \subset \widehat{\G}(\s)\,,  \quad\s = 0, \dots, N_c-1 
\e 
which is generated by the twisted infinite-dimensional subalgebra $\hat{h}(\s) \subset \hat{g}(\s)$.
According to (\ref{gmon}) and (\ref{[sm,sg]=0}), these subgroup orbifold elements satisfy
\begin{subequations}\label{hmono}
 \begin{gather}
 [\sm(\st,\s),\hat{h}(\st,\xi,t,\s)]  =0 \label{[sm,sh]=0}\, , \quad\quad \forall \; \hat{h}(\st, \xi, t, \s) \in \widehat{\HH}(\s) \\
\hat{h}(\st, \xi + 2\pi, t, \s)  = E(T, \s) \hat{h}(\st, \xi, t, \s) E(T, \s)^*\,  \label{hmonE}\\ 
\hat{h} (\st,\xi + 2\pi,t,\s)_{\Nrm}{}^{\Nsn}  =   e^{-2 \pi i
\frac{N(r)-N(s)}{R(\s)}} \hat{h} (\st,\xi,t,\s)_{\Nrm}{}^{\Nsn}\,  
\end{gather}
\end{subequations}
because each subgroup orbifold element is also a group orbifold element. It should be emphasized however that the monodromy of $\hat{h} \equiv \hat{h}(\st, \xi, t,\s)$ in (\ref{hmonE}) can look quite different when the support of $\hat{h}$ is taken into account (see the examples in Secs.\,\ref{sect:copy}, \ref{sect:diag} and \ref{sect:other_ex}). 

Using the properties of the twisted data matrix in Eqs.\,(\ref{[sm,sg]=0}) and (\ref{[sm,sh]=0}), and following steps which are analogous to those employed in the untwisted case, we may derive the \emph{twisted Polyakov-Wiegmann identity}
\begin{subequations}\label{twistedPW}
\begin{gather}
\widehat{S}_{\hg(\s)}[\M,\hg\hat{h}]  = \widehat{S}_{\hg(\s)}[\M, \hg] + \widehat{S}_{\hg(\s)}[\M, \hat{h}] - \frac{1}{4\pi}\int d^2\xi \0b \widehat{Tr}\Big( \M \0b {\hg}^{-1} \pl_+ \hg \0b \pl_-\hat{h} \hat{h}^{-1}\Big)  \\
\M \!\equiv \!\M(\st(T\!,\s), \s), \quad\!\! \hat{g}\!\equiv\! \hat{g}(\st(T\!,\s), \xi, t,\s), \quad\!\! \hat{h} \!\equiv \!\hat{h} (\st(T\!,\s), \xi, t,\s), \; \;\forall\,T \; \text{for each } \s\! =\! 0,\! \dots,\! N_c\!-\!1\;  
\end{gather}
\end{subequations}
for each twisted subgroup $\widehat{\HH}(\s)$ in sector $\s$ of any WZW orbifold. Here $\widehat{S}_{\hg(\s)}[\M, \hg]$ is the general WZW orbifold action (\ref{gorbact}) and, as expected, the twisted Polyakov-Wiegmann identity (\ref{twistedPW}) reduces to the ordinary Polyakov-Wiegmann identity \cite{PolyaWieg84} for the symmetric WZW system (\ref{untw_WZW})-(\ref{215a}) in the untwisted sector $\s = 0$ of each WZW orbifold. 

Following the conventional strategy, we may then define an action 
\begin{subequations} \label{cosetaction}
\begin{gather}
\widehat{S}_{\hg(\s)/\hat{h}(\s)}[\M, \hat{g}, \hat{h}_+, \hat{h}_-]   \equiv  \; \widehat{S}_{\hg(\s)}[\M, \hat{h}_+ \hat{g} \hat{h}_-] - \widehat{S}_{\hg(\s)}[\M, \hat{h}_+ \hat{h}_-]\,,  \quad\s = 0, \dots, N_c-1 \\ 
 \forall \; \hg \in \widehat{\G}(\s), \quad \forall\; \hat{h}_\pm \in \widehat{\HH}(\s) \subset \widehat{\G}(\s)
\end{gather}
\end{subequations}
for arbitrary group orbifold elements $\hat{g}$ and arbitrary subgroup orbifold elements $\hat{h}_\pm$ (which also satisfy Eq. (\ref{hmono})). By construction, the action (\ref{cosetaction}) is invariant 
\be\label{ginv2}
\widehat{S}_{\hg(\s)/\hat{h}(\s)}[\M, \hat{g}^{\hat{\psi}}, \hat{h}_+^{\hat{\psi}}, \hat{h}_-^{\hat{\psi}}] = \widehat{S}_{\hg(\s)/\hat{h}(\s)}[\M, \hat{g}, \hat{h}_+, \hat{h}_-] ,  \quad\s = 0, \dots, N_c-1 
\e
under the twisted vector gauge transformation 
\begin{subequations}\label{twgauge_trafo}
\be
\hat{g} (\st,\xi, t, \s) &\rightarrow& \hat{g}(\st, \xi, t, \s)^{\hat{\psi}} \equiv \hat{\psi} (\st,\xi, t, \s) \hat{g} (\st,\xi, t, \s) \hat{\psi}^{-1}(\st,\xi, t, \s)\; \\
\hat{h}_+(\st, \xi, t, \s) &\rightarrow&\hat{h}_+(\st, \xi, t, \s)^{\hat{\psi}} \equiv \hat{h}_+(\st, \xi, t, \s) \hat{\psi}^{-1}(\st, \xi, t, \s)\;  \\
\hat{h}_-(\st, \xi, t, \s) &\rightarrow& \hat{h}_-(\st, \xi, t, \s)^{\hat{\psi}} \equiv  \hat{\psi}(\st, \xi, t, \s) \hat{h}_-(\st, \xi, t, \s)\;  
\e
\vspace{-0.3in}
\be
\hat{g}(\st, \xi, t,\s)^{\hat{\psi}} \in \widehat{\G}(\s) \, , \quad \hat{h}_{\pm}(\st, \xi, t,\s)^{\hat{\psi}} \in \widehat{\HH}(\s)\subset \widehat{\G}(\s) . 
\e
\end{subequations}
Here the gauge transformation matrix $\hat{\psi}(\st, \xi, t, \s)$ is any element of $\widehat{\HH}(\s)$, whose properties
\begin{subequations}\label{Lmono}
\begin{gather}
 [\sm(\st,\s),\hat{\psi}(\st,\xi,t,\s)]  =0 \label{[sm,sL]=0}\, , \quad\quad \forall\; \hat{\psi}(\st, \xi, t, \s) \in \widehat{\HH}(\s) \\
\hat{\psi}(\st, \xi + 2\pi, t, \s)= E(T, \sigma) \hat{\psi}(\st, \xi,t,\s) E(T,\s)^* \\
\hat{\psi} (\st, \xi+2\pi, t,\s)_{N(r) \mu}{}^{N(s) \nu} 
 = e^{-2\pi i \srac{N(r)-N(s)}{R(\sigma)}} \hat{\psi}(\st, \xi,t,\s)_{N(r) \mu}{}^{N(s) \nu}
\end{gather}
\end{subequations}
also follow because $\hat{\psi} \in \widehat{\HH}(\s) \subset \widehat{\G}(\s)$. It will be convenient to write the monodromies (\ref{gmon}), (\ref{hmono}) and (\ref{Lmono}) in the unified notation
\be
\hat{O}(\st, \xi+2\pi, t, \s) = E(T, \s) \hat{O}(\st, \xi, t, \s) E(T, \s)^*\, , \qquad \hat{O} = \hat{g}, \hat{h}_{\pm} \text{ or } \hat{\psi}.
\e 
Then we verify the combined gauge/monodromy relations 
\begin{subequations}\label{gaugedmono}
\begin{align}
\hat{O}(\st, \xi+2\pi, t, \s)^{\hat{\psi}} & = \hat{\psi}(\st, \xi+2\pi, t, \s) \hat{O}(\st, \xi+2\pi, t, \s) \hat{\psi}^{-1}(\st, \xi+2\pi, t, \s) \\
&=  E(T, \s)\hat{\psi}(\st, \xi, t, \s) \hat{O}(\st, \xi, t, \s) \hat{\psi}^{-1}(\st, \xi, t, \s) E(T, \s)^* \\ &= 
 E(T, \s) \hat{O}(\st, \xi, t, \s)^{\hat{\psi}} E(T, \s)^*\,, \quad\quad \hat{O} = \hat{g} \text{ or } \hat{h}_{\pm}
\end{align}
\end{subequations}
which tell us that the monodromies of these fields are preserved by the twisted gauge transformation. Alternately, (\ref{gaugedmono}) 
says that the twisted gauge transformation commutes with the monodromy transformations. 


Using the twisted Polyakov-Wiegmann identity (\ref{twistedPW}) we may re-express the action (\ref{cosetaction}) as the {\em general  coset orbifold action}
\begin{subequations}\label{empty_+} 
\begin{align} \label{empty}
\widehat{S}_{\hg(\s)/\hat{h}(\s)}[\M, \hg, \hat{A}_{\pm}] =  \\
 = \widehat{S}_{\hg(\s)}[\M, \hg] \; +& \frac{1}{4\pi}\int d^2 \xi \0b \widehat{Tr} \Big(\M \big(\0b\hat{g}^{-1} \pl_+ \hg\0b (i \hat{A}_-) - i \hat{A}_+ \0b \pl_-\hg \hat{g}^{-1} - \hat{g}^{-1} \hat{A}_+ \hg \hat{A}_- + \hat{A}_+ \hat{A}_-\big)\Big)\, \nn
\end{align}
\begin{gather}
\M \!\equiv \!\M(\st(T\!,\s), \s), \quad\!\! \hat{g}\!\equiv\! \hat{g}(\st(T\!,\s), \xi, t,\s), \quad\!\! \hat{A}_\pm \!\!\equiv \!\hat{A}_\pm (\st(T\!,\s), \xi, t,\s), \; \;\forall\,T \; \text{for each } \s\! =\! 0,\! \dots,\! N_c\!-\!1 \\ 
\hat{A}_+ (\st\!, \xi, t, \s) \equiv\! - i \hat{h}_+^{-1}(\st\!, \xi, t, \s)\partial_+ \hat{h}_+(\st\!, \xi, t, \s)\, , \; \hat{A}_- (\st\!, \xi, t, \s) \equiv\! - i  \hat{h}_-(\st\!, \xi, t, \s) \partial_-\hat{h}_-^{-1}(\st\!, \xi, t, \s)\,  \label{Adefi}\\
\hat{A}_\pm (\st, \xi, t, \s) \in \hat{h}(\s) \subset \hg(\s)
\end{gather}
\end{subequations}
where $\widehat{S}_{\hg(\s)}[\M, \hg]$ is the general WZW orbifold action (\ref{gorbact}) and  $\{\hat{A}_{\pm}\}$ are the {\em twisted matrix gauge fields}. The gauged WZW orbifold action (\ref{empty_+}) is the action formulation in sector $\s$ of the general coset orbifold $A_{g/h}(H)/H$.

\section{Properties of the General Coset Orbifold Action}

We now discuss a number of properties of the general coset orbifold action (\ref{empty_+}), beginning with the subject of {\em monodromy}. The monodromy of the group orbifold elements $\hg$ is given in Eq. (\ref{gmon}) and,
according to (\ref{hmono}) and (\ref{Adefi}), the twisted matrix gauge fields satisfy 
\begin{subequations}\label{A_mono}
\begin{gather}
\hat{A}_{\pm}(\st, \xi + 2\pi, t, \s)  = E(T, \sigma) \hat{A}_{\pm}(\st, \xi, t,\s) E(T,\s)^* \; \label{A_mono_a}\\
  \hat{A}_{\pm} (\st, \xi+2\pi, t, \s)_{N(r) \mu}{}^{N(s) \nu} 
= e^{-2\pi i \srac{N(r)-N(s)}{R(\sigma)}} \hat{A}_{\pm}(\st, \xi, t,\s)_{N(r) \mu}{}^{N(s) \nu} \,\\
[\sm(\st,\s),\hat{A}_\pm(\st,\xi,t,\s)]  =0 \label{[sm,sA]=0}\, .
\end{gather}
\end{subequations}
Then we verify that the general coset orbifold action (\ref{empty_+}) has trivial monodromy 
\begin{gather}
\widehat{S}_{\hg(\s)/\hat{h}(\s)}[\M, \hg (\xi +2\pi), \hat{A}_{\pm}(\xi + 2\pi)]  = \widehat{S}_{\hg(\s)/\hat{h}(\s)}[\M, \hg(\xi), \hat{A}_{\pm}(\xi)],  \quad\s = 0, \dots, N_c-1 \;  
\end{gather}
according to the ${\cal M}$-selection rule (\ref{eq:msel}).

Turning now to the subject of {\em gauge transformations}, we see that the twisted matrix gauge fields (\ref{Adefi}) transform under the twisted gauge transformations (\ref{twgauge_trafo}) as
\begin{subequations}\label{gauged_A}
\begin{gather}
\hat{A}_{\pm}(\st, \xi, t, \s) \rightarrow  \hat{A}_{\pm}(\st, \xi, t, \s)^{\hat{\psi}} \, \\
\hat{A}_{\pm}(\st, \xi, t, \s)^{\hat{\psi}} = 
\hat{\psi}(\st, \xi, t, \s) \hat{A}_\pm(\st, \xi, t, \s) \hat{\psi}^{-1}(\st, \xi, t, \s) + i \pl_\pm\hat{\psi}(\st, \xi, t, \s) \, \hat{\psi}^{-1}(\st, \xi, t, \s) \, .
\end{gather}
\end{subequations}
Then, using (\ref{A_mono_a}) and (\ref{gauged_A}), one finds the combined gauge/monodromy relations 
\be
\hat{O}(\st, \xi+2\pi, t, \s)^{\hat{\psi}} =  E(T, \s) \hat{O}(\st, \xi, t, \s)^{\hat{\psi}} E(T, \s)^*\, , \quad\quad \hat{O} = \hat{g},  \hat{h}_{\pm}  \text{ or } \hat{A}_\pm\,  
\e
which generalize Eq. (\ref{gaugedmono}). The twisted gauge invariance of the general coset orbifold action
\be
\widehat{S}_{\hg(\s)/\hat{h}(\s)}[\M, \hg^{\hat{\psi}}, \hat{A}_{\pm}^{\hat{\psi}}] = \widehat{S}_{\hg(\s)/\hat{h}(\s)}[\M, \hg, \hat{A}_{\pm}] ,  \quad\s = 0, \dots, N_c-1
\e
is another form of Eq. (\ref{ginv2}).


We turn next to the {\em equations of motion} of the general coset orbifold action. Because the twisted data matrix ${\cal M}$ is invertible and commutes with $\hg$ and $\hat{A}_\pm$, one obtains the twisted equations of motion of sector $\s$ of $A_{g/h}(H)/H$ 
\begin{subequations}\label{EOMtw}
\begin{gather}
\Big(\widehat{D}_-(\hg^{-1}\widehat{D}_+\hg) + i \widehat{F}_{-+}\Big)_{\hat{g}(\s)} = 0\,, \qquad  \Big(\hg^{-1} \widehat{D}_+\hg \Big)_{\hat{h}(\s)} = \Big((\widehat{D}_-\hg)\hg^{-1}\Big)_{\hat{h}(\s)} = 0\\
\widehat{D}_\pm(\cdot) = \partial_{\pm}(\,\cdot\,) + i [\hat{A}_\pm, \,\cdot \,],  \quad \widehat{F}_{-+}\equiv \partial_-\hat{A}_+ - \partial_+ \hat{A}_- + i [\hat{A}_-, \hat{A}_+]
\end{gather}
\end{subequations}
by varying the general coset orbifold action (\ref{empty_+}). 
 Here $(\cdot)_{\hat{g}(\s)}$ or $(\cdot)_{\hat{h}(\s)}$ means that the expression $(\cdot)$ is valued respectively on the twisted ambient algebra $\hat{g}(\s)$ or the twisted subalgebra $\hat{h}(\s)$. The twisted covariant derivatives $\widehat{D}_\pm$ preserve the monodromies and the twisted gauge transformation properties of the group orbifold elements $\hat{g}$ and the subgroup orbifold elements $\hat{h}$
\begin{subequations}
\begin{gather} 
\Big(\widehat{D}_\pm \hat{O}\Big)(\st, \xi+2\pi, t, \s) = E(T, \s) \Big(\widehat{D}_\pm \hat{O}(\st, \xi, t, \s)\Big) E(T, \s)^*\\
\Big(\widehat{D}_\pm \hat{O}(\st, \xi, t, \s) \Big)^{\hat{\psi}} = \hat{\psi}(\st, \xi, t, \s) \widehat{D}_\pm \hat{O}(\st, \xi, t, \s) \hat{\psi}^{-1}(\st, \xi, t, \s), 
\qquad \hat{O} = \hat{g} \text{ or } \hat{h}_{\pm} 
\end{gather}
\end{subequations}
and the twisted field strength $\widehat{F}_{-+}$ also satisfies the relations
\begin{subequations}
\begin{gather}
\widehat{F}_{-+}(\st, \xi+2\pi, t, \s) = E(T, \s) \widehat{F}_{-+}(\st, \xi, t, \s) E(T, \s)^*\\
\widehat{F}_{-+}(\st, \xi, t, \s)^{\hat{\psi}} = \hat{\psi}(\st, \xi, t, \s) \widehat{F}_{-+}(\st, \xi, t, \s) \hat{\psi}^{-1}(\st, \xi, t, \s). 
\end{gather}
\end{subequations}
It follows that the twisted equations of motion (\ref{EOMtw}) are covariant both under monodromy transformations and twisted gauge transformations.
An equivalent form of the equations of motion 
\begin{subequations}
\begin{gather}
\widehat{F}_{-+} = 0\, \quad \longleftrightarrow \quad [\widehat{D}_-(\cdot), \widehat{D}_+(\cdot)] = 0 \\
\widehat{D}_-(\hg^{-1}\widehat{D}_+\hg)=\widehat{D}_+((\widehat{D}_- \hg) \hg^{-1}) = 0\,,  \qquad  \Big(\hg^{-1} \widehat{D}_+\hg \Big)_{\hat{h}(\s)} = \Big((\widehat{D}_-\hg)\hg^{-1}\Big)_{\hat{h}(\s)} = 0
\end{gather}
\end{subequations}
also follows because $\widehat{F}_{-+}$ is valued on $\hat{h}(\s)$. For brevity we refrain here from using the equations of motion to integrate out the twisted matrix gauge fields, which would give the twisted sigma model form of the general coset orbifold action. 

We finally discuss the {\em untwisted sector} $\s=0$ of the orbifold $A_{g/h}(H)/H$.  In this case the general coset orbifold action (\ref{empty_+}) reduces to the ordinary gauged WZW  action [18--22] 
for the special case of the $H$-invariant coset construction $A_{g/h}(H)$. Except for the presence of the untwisted data matrix $M=M(k,T)$, the form of this action is well known:
\begin{subequations} \label{old_empty}
\begin{align}
\frac{g}{h} = \frac{g}{h}(H) \equiv \frac{\hat{g}(0)}{\hat{h}(0)}   \; , \quad\quad h_{\pm}(T, \xi, t) \equiv \hat{h}_{\pm} (\st, \xi, t, 0) \;,  \quad A_\pm(T,\xi, t) \equiv \hat{A}_{\pm}(\st, \xi, t, 0) \in h
\end{align}
\begin{align}
S_{g/h}[M, g, A_{\pm}] = \quad & \\
\quad = S_{WZW}[M, g] \;& + \frac{1}{4\pi}\int d^2 \xi \0b Tr \Big(M \big(\0b g^{-1} \pl_+ g\0b (i A_-) - i A_+ \0b \pl_- g g^{-1} - g^{-1} A_+ g A_- + A_+ A_-\big)\Big)\, \nn
\end{align}
\be
A_+ (T, \xi, t) \equiv - i h_+{}^{-1}(T, \xi, t)\partial_+ h_+(T, \xi, t)\, , \quad A_- (T, \xi, t) \equiv - i h_-(T, \xi, t) \partial_-h_-^{-1}(T, \xi, t)\,  \label{old_Adefi}
\e
\begin{gather}
[M(k,T), g(T, \xi, t)] = [M(k,T), h_{\pm}(T, \xi, t)] = [M(k,T), A_{\pm}(T, \xi, t)]= 0.\label{morecom}
\end{gather}
\end{subequations}
The commutator $[M, g]=0$ in (\ref{morecom}) was discussed in Sec.\,\ref{sect:action}, and the other commutators in (\ref{morecom}) follow because $h(T, \xi, t)$ is an untwisted subgroup element.
This action has the expected untwisted gauge invariance
\begin{subequations}\label{obv1}
\begin{gather}
 \psi(T, \xi, t) \equiv \hat{\psi}(\st, \xi, t, 0), \qquad
[M(k,T), \psi(T, \xi, t) ] =0 \label{obvious1} 
\end{gather}
\vspace{-.4in}
\begin{eqnarray}
g (T,\xi, t) &\rightarrow& {g}(T, \xi, t)^\psi \equiv \psi (T,\xi, t) g(T,\xi, t) {\psi}^{-1}(T,\xi, t)\; \\
{h}_+(T, \xi, t) &\rightarrow& {h}_+(T, \xi, t)^\psi \equiv {h}_+(T, \xi, t) {\psi}^{-1}(T, \xi, t)\;  \\
h_-(T, \xi, t) &\rightarrow & h_-(T, \xi, t)^\psi \equiv  \psi(T, \xi, t) h_-(T, \xi, t)
\end{eqnarray}
\vspace{-.4in}
\begin{gather}
{A}_{\pm}(T, \xi, t)^\psi = 
{\psi}(T, \xi, t) {A}_\pm(T, \xi, t){\psi}^{-1}(T, \xi, t) + i \pl_\pm{\psi}(T, \xi, t, \s) \, {\psi}^{-1}(T, \xi, t) \\
 S_{g/h}[M, g^\psi, A_\pm^\psi] = S_{g/h}[M,g,A_\pm]
\end{gather}
\end{subequations}
where $\psi$ is the untwisted gauge transformation matrix. The gauge invariance (\ref{obvious1}) of $M$ follows because $\psi$ is in the subgroup.

Most important is the behavior of the symmetric theory under the symmetry group $H$. In particular, the untwisted quantities in (\ref{old_empty})-(\ref{obv1}) transform under $H$ as 
\begin{subequations}\label{O_transform}
\begin{gather}
O(T, \xi, t)' = W(\hsig; T) O(T, \xi, t) W^\dagger(\hsig ; T)\, , \quad O = g,  h_{\pm}, A_{\pm}   \text{ or } \psi \\
g(T, \xi, t)' \in \G, \quad h(T, \xi, t)' \in \HH, \quad A_\pm(T,\xi, t)' \in h, \quad \psi(T, \xi, t)' \in \HH\label{group_inv}. 
\end{gather}
\end{subequations}
where $W(\hsig;T) \in H$, seen earlier in Eqs.\,(\ref{H_sym})-(\ref{155}), is the action of $\hsig \in H$ in representation $T$ of Lie $g$. The relations in (\ref{O_transform}) guarantee that $H$ is an automorphism group both of the  algebra $g$ and subalgebra $h \subset g$. This means in particular that $h$ is an {\em $H$-covariant subalgebra of $g$} and that $g/h$ is an {\em $H$-invariant coset construction}.  The general $H$-invariant coset construction is discussed at the operator level in Refs.\,\cite{Evslin:1999ve,Halpern:2000vj}.
The final ingredient is the $H$-symmetry of the untwisted data matrix
\be
 W^\hcj(h_\s;T)M(k,T)W(h_\s;T) = M(k,T),\quad \forall \ h_\s \in H, \quad H
  \subset {\rm Aut}(g),  \quad H
  \subset {\rm Aut}(h) \label{H_sym2}
\e
and then one finds from (\ref{O_transform}) and (\ref{H_sym2}) that the coset action $S_{g/h}$ is $H$-invariant 
\begin{gather}
 S_{g/h}[M, g', A_\pm'] = S_{g/h}[M,g,A_\pm] \, .
\end{gather}
We also note the combined relations
\be
\big(O(T,\xi, t)^\psi\big)' = \big(O(T,\xi, t)'\big)^{\psi'} = W(\hsig; T) O(T, \xi, t)^\psi W^\dagger(\hsig; T), \quad O = g, h_{\pm} \text{ or } A_\pm
\e
which describe the exchange of order of the gauge and symmetry transformations.

The general coset orbifold action (\ref{empty_+}) can also be derived  from the untwisted $H$-invariant system  (\ref{old_empty}), (\ref{H_sym2}) by the method of eigenfields and local isomorphisms:
\begin{subequations}
\begin{gather}
O = g, h_{\pm}, A_\pm \text{ or } \psi, \\
{{\EuScript O}}(T,\xi, t) \equiv U(T, \s) { O} (T, \xi, t) U^\dagger(T, \s)\, , \quad {\EuScript O} = \sg, \shh_{\pm}, {\EuScript A}_\pm  \text{ or }{\Psi} \\
{{\EuScript O}}(T,\xi, t)' = E(T, \s) {\EuScript O} (T, \xi, t) E(T, \s)^*\\
{\cal M}(\st(T,\s), \s) = U(T, \s) M(k,T) U^\dagger(T, \s) \label{MMM}\\
{\EuScript O} \dual \hat{O}, \quad \hat{O} = \hg, \hat{h}_{\pm}, \hat{A}_{\pm}  \text{ or } \hat{\psi}\\
\text{automorphic response } E(T, \s) \dual \text{monodromy } E(T, \s) \\
\hat{O}(\st, \xi+2\pi, t, \s) = E(T, \s) \hat{O}(\st, \xi, t, \s) E(T, \s)^*\,    
\end{gather}
\end{subequations}
where ${\cal O}$ are the eigenfields and  $\{U^\dagger(T,\s), E(T, \s)\}$ is the eigendata of the extended $H$-eigenvalue problem in Eq. (\ref{exH}).

\section{Twisted Algebras and Subalgebras: The Recipe}\label{sect:twistedsubal}

So far, we have discussed only the features of the general coset orbifold action (\ref{empty_+}) which hold for all coset orbifolds. To find the appropriate set of coset orbifold actions for any particular coset orbifold, we need to know the embedding of the twisted subgroup $\widehat{\HH}(\s)  \subset \widehat{\G}(\s)$, $\srange$ at the action level for each sector of that orbifold. 
In this section we extend the review of  Sec.\,\ref{sectbt} to discuss the twisted affine embeddings $\{\hhs \subset \hgs$, $\s =0, \dots, N_c-1\}$ of the coset orbifold \cite{Evslin:1999ve,Halpern:2000vj} at the operator level, and we will also use the twisted  affine embeddings to solve the corresponding embedding problem at the action level. 


We begin with the ambient affine algebra $\hgs$.
The twisted current-current OPE's  \cite{deBoer:1999na, Halpern:2000vj, deBoer:MBH01} and the mode expansion of the twisted $g$ currents $\hj_{\hgs}$ in (\ref{current_components}) 
\be
\hj_{n(r)\mu}(z,\s) = \sum_{m \in \Z} \hj_{n(r)\mu} \big(m + \srac{n(r)}{\rho(\s)}\Big) z^{- \big(m + \frac{n(r)}{\r(\s)}\big) -1}
\e
give the {\em general twisted current algebra} $\hgs$ of sector $\s$ of the WZW orbifold $A_g(H)/H$ 
\begin{gather}
 [\hj_\nrm(\mnrrs),\hj_\nsn(\nnsrs)]=i\scf_{\nrm;\nsn}{}^{n(r)+n(s),\delta}(\s)
 \hj_{n(r)+n(s),\delta}(\mnnrnsrs) \nn \\
\quad\quad\quad\quad\quad\quad\quad\quad\quad\quad\quad\quad\quad\quad + (\mnrrs)\delta_{\mnnrnsrsf,0}\sG_{\nrm;\mnrn}(\s)\, \label{mcall}\\
m,n \in \Z, \quad \s = 0, \dots, N_c-1 .\nn
\end{gather}
Here $\sG(\s)$ and $\scf(\s)$ are called the {\em twisted metric} and {\em twisted structure constants} of $\hgs$, explicit formulae for which are given in Refs.\,\cite{deBoer:1999na, Halpern:2000vj, deBoer:MBH01}. It will be helpful to write the algebra $\hgs$ symbolically as 
\begin{gather}\label{h_symbol}
[\hj_{\hgs}(\cdot), \hj_{\hgs}(\cdot) ] = i {\cal F}_{\hgs} \hj_{\hgs}(\cdot) + \dots
\end{gather}
where ${\cal F}_{\hgs} \equiv \{ {\cal F}(\s) \}$. 

Next, for each particular coset orbifold $A_{g/h}(H)/H$  one needs to identify the twisted affine subalgebra $\hat{\h}(\s) \subset \hat{\g}(\s)$  in each sector $\s$  of the orbifold. This problem was studied in Refs.\,\cite{Evslin:1999ve,Halpern:2000vj}. In summary, one uses local isomorphisms (see Eq.\,(\ref{lim})) to find the {\em affine embedding matrix} ${\cal E}(\s)$ of the twisted $h$ currents $\hat{J}_\hhs$ in the twisted $g$ currents $\hat{J}_\hgs$
\begin{subequations}\label{current_embedding}
\be
J_h(\cdot) = {\cal E}(0) J_g(\cdot) \dual \hat{J}_\hhs(\cdot) = {\cal E}(\s) \hat{J}_\hgs (\cdot)
\e
\begin{gather}\label{affine_sub}
[\hj_{\hhs}(\cdot), \hj_{\hhs}(\cdot) ] = i {\cal F}_{\hhs} \hj_{\hhs}(\cdot) + \dots\end{gather}
\end{subequations}
given the affine embedding matrix ${\cal E}(0)$ in the untwisted sector $\s=0$ of the orbifold. The twisted structure constants  ${\cal F}_{\hhs}$ are  the restriction of the twisted structure constants  ${\cal F}_{\hgs}$ to the twisted affine subalgebra $\hhs$. The affine embedding matrix ${\cal E}(\s), \s =0, \dots, N_c-1$ was constructed for a large class of cyclic coset orbifolds in Ref.\,\cite{Evslin:1999ve}, and we will discuss this class of examples and other examples below. Our next task is to translate the affine embedding $\hhs\subset \hgs$  into its action analogue $\hat{h}(\s)\subset \hg(\s)$. 

The key to this translation is the set of {\em twisted representation matrices} $\{\st(T,\s)\}$ of sector $\s$ of $A_g(H)/H$, whose general form is \cite{deBoer:MBH01}
\begin{subequations}\label{65}
\begin{gather}
T\dual \st(T,\s)\\
\st_{n(r)\mu}(T,\s) = \chi_{n(r)\mu}(\s) U(\s)_{n(r)\mu}{}^a U(T,\s) T_a U^\dagger(T,\s), \quad \forall\, T \quad \text{for each } \s = 0, \dots, N_c-1
\end{gather}
\end{subequations}
where $U^\dagger(\s)$ and $U^{\dagger}(T,\s)$  are respectively the eigenvector matrices of the $H$-eigenvalue problem and the extended $H$-eigenvalue problem (\ref{exH}), while $\chi(\s)$ is a set of normalization constants. In this formula the matrices $\{T\}$ include all representations of Lie $g$, while the information about the symmetry group $H$ is encoded in the eigendata of the two eigenvalue problems. In current-algebraic orbifold theory, relations such as (\ref{MMM}) and (\ref{65}) - as well as the explicit forms of ${\cal L}_{\hgs}(\s)$, $\sG(\s)$ and $\scf(\s)$ in Refs.\, \cite{deBoer:1999na, Halpern:2000vj, deBoer:MBH01} - are called {\em duality transformations}, which are discrete Fourier transforms of various quantities in the symmetric theory. The explicit form of $\{\st(T,\s)\}$ for the WZW permutation orbifolds is reviewed in Sec.\,\ref{sectperm}.

Further properties of the twisted representation matrices are as follows \cite{deBoer:MBH01}.

In the first place, the twisted representation matrices appear in the {\em tangent space description} of the group orbifold elements
\begin{subequations}
\begin{gather}\label{tw_expon}
\hat{g} (\st(T,\s),\xi,t,\s) =   e^{i \hat{\beta}^{\nrm} (\xi,t,\s) \st_{\nrm} (T,\s)} \equiv e^{i \hat{\beta}_{\hat{g}(\s)}(\xi, t,\s) \cdot \st_{g(\s)}(T,\s)} \in \widehat{\G}(\s)\\
 \forall \hat{\beta},\; \forall\, T \quad \text{for each } \s = 0, \dots, N_c-1
\end{gather}
\end{subequations}
where the quantities $\{\hat{\beta}\}$ in (\ref{tw_expon}) are called the {\em twisted tangent space coordinates} with definite monodromy 
\be\label{b-mono}
\hat{\beta}^{\nrm}(\xi + 2\pi, t, \s) =\hat{\beta}^{\nrm}(\xi, t, \s)E_{n(r)}(\s)^*= \hat{\beta}^{\nrm}(\xi, t, \s)  e^{2\pi i \frac{n(r)}{\rho(\s)}}\,,  \quad\s = 0, \dots, N_c-1 . 
\e
Then the selection rule \cite{deBoer:1999na, Halpern:2000vj, deBoer:MBH01} for the twisted representation matrices 
\begin{subequations}
\label{sel-rule-for-st-1}
\begin{gather}
 \st_\nrm(T,\s) = E_{n(r)}(\s) \big{(}\,E(T,\s)
 \st_\nrm(T,\s) E(T,\s)^\ast\,\big{)} \quad \quad\quad\quad\quad \quad \\ \Leftrightarrow \quad \quad \quad
 \st_\nrm(T,\s)_\Nsn{}^\Ntd ( 1- e^{-2\pi i(\frac{n(r)}{\r(\s)}+
 \frac{N(s)-N(t)}{R(\s)}  )}) = 0
\end{gather}
\end{subequations}
guarantees the consistency  of the monodromy of $\{\hat{\beta}\}$ in (\ref{b-mono}) with the monodromy (\ref{gmon}) of the group orbifold elements. Similarly, the selection rule (\ref{sel-rule-for-st-1}) guarantees that the monodromy of the classical matrix current $\hj(\st)$ in (\ref{EOMtwis})
\be\label{jcomp}
\hj(\st\!, \xi, t, \s) =  -\frac{i}{2}\hat{g}^{-1}(\st\!,\xi,t,\s)\pl_+
  \hat{g}(\st\!,\xi,t,\s)=
\hj_{\nrm}(\xi, t, \s) \sG^{\nrm; \mnrn}(\s) \st_{\mnrn}(T,\s)
\e
is consistent with the monodromy of the current components $\hj_{\nrm}$ in (\ref{current_components}). (The twisted tensor ${\cal G}^{\bullet}(\s)$ in (\ref{jcomp}) is called the inverse twisted metric of sector $\s$.)

The twisted representation matrices also satisfy the {\em orbifold Lie algebra} $g(\s)$
\begin{subequations}\label{eq:reason-for-casimir-of-st}
\begin{equation}
 \left[ \st_\nrm(T,\s) ,\st_\nsn(T,\s) \right] =
 i\scf_{\nrm;\nsn}{}^{n(r)+n(s),\d}(\s) \st_{n(r)+n(s),\d}(T,\s) \, 
\end{equation}
\begin{gather}
[\st_{g(\s)}(T,\s), \st_{g(\s)}(T,\s) ] = i {\cal F}_{\hgs} \st_{g(\s)}(T,\s)\, , \quad \st_{g(\s)}(T,\s) \in g(\s), \quad \forall\,T \quad \text{for each }\s = 0, \dots, N_c-1 
\end{gather}
\end{subequations}
copies of which form the twisted infinite-dimensional Lie algebra
\be
\hat{g}(\s) = \bigoplus_\xi g_\xi(\s), \quad g_\xi(\s) \cong g(\s) \,,  \quad\s = 0, \dots, N_c-1 
\e
which was used to label the general WZW orbifold action $\widehat{S}_{\hat{g}(\s)}$. Most important, the twisted structure constants ${\cal F}_{\hgs}= \{{\cal F}(\s)\}$ of the orbifold Lie algebra $g(\s)$ are the {\em same} as the twisted structure constants of the general twisted current algebra $\hgs$ in (\ref{mcall}).

Since the algebras (\ref{mcall}) and (\ref{eq:reason-for-casimir-of-st}) share the same twisted structure constants ${\cal F}_{\hgs}$, the twisted representation matrices $\{\st_{h(\s)}\}$ which generate the {\em orbifold Lie subalgebra} ${h}(\s) \subset g(\s)$ are immediately identified as 
\begin{subequations}
\begin{gather}
\st_{{h}(\s)}(T,\s) \equiv {\cal E}(\s) \st_{g(\s)}(T,\s)\,  \; \in {h}(\s) \subset {g}(\s), \quad \forall\,T \quad \text{for each }\s = 0, \dots, N_c-1 \\
[\st_{{h}(\s)}(T,\s), \st_{{h}(\s)}(T,\s)] = i {\cal F}_{\hhs} \st_{{h}(\s)}(T,\s)\label{tw_sub_alg}\quad \quad 
\hat{h}(\s) = \bigoplus_\xi h_\xi(\s), \quad h_\xi(\s) \cong h(\s)
\end{gather}
\end{subequations}
where ${\cal E}(\s)$ is the affine embedding matrix in (\ref{current_embedding}). By construction, the structure constants ${\cal F}_{\hhs}$ of the orbifold Lie subalgebra ${h}(\s)$ in (\ref{tw_sub_alg}) are the same as those of  the twisted affine subalgebra $\hhs$ in (\ref{affine_sub}). It follows that the tangent space description of the subgroup orbifold elements  $\hat{h}$ is
\begin{subequations}\label{613}
\begin{gather}
\hat{h}(\st(T,\s), \xi, t, \s) = e^{i \hat{\beta}_{\hat{h}(\s)}(\xi, t, \s) \cdot \st_{{h}(\s)}(T,\s)} =
e^{i\{ \hat{\beta}_{\hat{h}(\s)}(\xi, t, \s)\cdot{\cal E}(\s) \} \cdot \st_{{g}(\s)}(T,\s)} \in \widehat{\HH}(\s) \subset \widehat{\G}(\s) \\
\forall \; \hat{\beta}_{\hat{h}(\s)}, \; \forall \, T \quad  \text{for each }  \s = 0, \dots, N_c-1.
\end{gather}
\end{subequations}
The quantities $\{\hat{\beta}_{\hat{h}(\s)}(\xi, t,\s)\cdot{\cal E}(\s)\}$ in (\ref{613}) are the restriction to $\hat{h}(\s) \subset \hat{g}(\s)$ of the twisted tangent space coordinates $\{\hat{\beta}\}$, which guarantees the monodromy of the subgroup orbifold element $\hat{h}(\st, \xi, t,\s)$ in  (\ref{hmono}). Finally, the tangent space description (\ref{613}) of the subgroup orbifold elements provides a complete description of the matrix gauge fields $\hat{A}_\pm$ in the coset orbifold action (\ref{empty_+}). 

In this section, we have provided a simple {\em recipe}
\begin{gather}\label{recipe}
\hj_{\hhs} = {\cal E}(\s) \hj_{\hgs} \longrightarrow \st_{h(\s)}(T,\s)= {\cal E}(\s) \st_{g(\s)}(T,\s), \quad \forall\,T \quad \text{for each }
 \srange
\end{gather}
which uses the solution of the operator embedding problem to find the appropriate embedding of the twisted subgroup $\widehat{\HH}(\s) \subset \widehat{\G}(\s)$ in each sector $\s$ of any particular coset orbifold $A_{g/h}(H)/H$. This completes our  discussion of the general coset orbifold and we turn now to some large examples, including in particular the general permutation coset orbifold in Sec.\,\ref{sect:permcoset}. 


\section{Background: The WZW Permutation Orbifolds}\label{sectperm}

As the building blocks of permutation coset orbifolds,
we review here some salient features of the WZW permutation orbifolds \cite{deBoer:MBH01, MBHObers}
\begin{subequations}
\begin{gather}
\frac{A_g(H)}{H} \\
g = \bigoplus_{I=0}^{K-1} \gfrak^I, \quad \gfrak^I \cong \gfrak, \quad H(\text{permutation}) \subset S_N, \quad H(\text{permutation}) \subset Aut(g) \,\\
T = \bigoplus_{I=0}^{K-1} T^I, \quad T^I_a \cong T_a, \quad a = 1, \dots, \dim \gfrak
\end{gather}
\end{subequations} 
where $K \le N$, $\{T_a\}$ is an irreducible representation of $\gfrak$, and the permutations act among the copies of affine $\gfrak$ at level $k_I = k$.

For the twisted sectors of the WZW permutation orbifolds, one knows that \cite{Halpern:2000vj, deBoer:MBH01} 
\begin{subequations}
\begin{gather}
N(r) = n(r), \quad R(\s) = \rho(\s) \\
\hj_{n(r)\mu}(\s) \rightarrow \hj_{n(r) aj}(\s), \quad \hat{g}(\st(T,\s), \s)_{N(r) \mu}{}^{N(s)\nu} \rightarrow \hat{g}(\st(T,\s), \s)_{n(r) \alpha j}{}^{n(s) \beta l}\\
a= 1, \dots, \dim \gfrak, \quad \alpha, \beta = 1, \dots, \dim T ,  \quad\s = 0, \dots, N_c-1
\end{gather}
\end{subequations}
where $\dim T$ is the size of the $\g$ irreps $\{ T_a\}$.
Moreover, we will follow the relabelling introduced in Ref.\,\cite{MBHObers}, where the relations \cite{deBoer:MBH01} 
\be\label{permlabel}
\frac{N(r)}{R(\s)}=\frac{n(r)}{\r(\s)} = \frac{\hjs}{f_j(\s)},  \quad \bar{\hat{j}} =0, \dots, f_j(\s)-1, \quad \sum_j f_j(\s) = K \le N
\e
were used to eliminate the spectral integers $n(r)$ in favor of $\hat{j}$, for example
\begin{subequations}
\begin{gather}
\hj_{n(r) a j}(\s) \rightarrow \hj_{\hat{j} a j}(\s), \quad \hat{g}(\st(T,\s), \s)_{n(r) \alpha j}{}^{n(s) \beta l} \rightarrow \hat{g}(\st(T,\s), \s)_{\hat{j} \alpha j}{}^{\hat{l} \beta l}, \quad E_{n(r)}(\s) \rightarrow e^{-2\pi i \srac{\hat{j}}{f_j(\s)}} \\
\hj_{\hat{j} a j}( z e^{2\pi i }, \s) = e^{-2\pi i \frac{\hat{j}}{f_j(\s)}} \hj_{\hat{j} aj}(z, \s), \quad \hj_{\hat{j}a j}(z,\s)  = \sum_{m \in \Z} \hj_{\hat{j} a j}(m + \srac{\hat{j}}{f_j(\s)}) z^{-(m + \srac{\hat{j}}{f_j(\s)}) -1}.
\end{gather}
\end{subequations}
We remind the reader that, in this notation, each element $\hsig \in H(\text{permutation})$ has been expressed in terms of disjoint cycles $j$ of size and order $f_j(\s)$, where $\hat{j}$ runs within each disjoint cycle. As a computational aid to the reader, we note the following special cases: 
\begin{subequations}
\begin{eqnarray}
 \Z_\l \!\!\!\!\!&:& \!\!
 K = \lambda, \quad \!\! f_j(\s)\! = \!\rho(\s), \quad\!\!\! \bar{\hat{j}} = 0,..., \rho(\s)\! -\!1, \quad\!\! j = 0, ..., \frac{\lambda}{\rho(\s)}\! -\!1,  \; \s = 0,..., \rho(\s)\!-\!1 \label{62a} \\
 \Z_\l,& &\!\!\!\!\!\!\!\!\! \l = \text{prime}: \quad   
\r(\s) = \lambda, \quad  \bar{\hat{j}}\! =\! 0,\dots, \l \!-\!1, \quad  j\!=\!0, \quad \s =1, \dots, \l -1 \, \\ 
 S_N \!\!\!\!\!&:&\!  
 K=N, \quad \! f_j(\s) = \s_j, \quad \s_{j+1} \leq \s_j, \; j = 0, \dots, n(\vec{\s})-1, \quad\sum_{j=0}^{n(\vec{\s})-1} \s_j = N .\label{62c}
\end{eqnarray}
\end{subequations}
Here $\r(\s)$ in (\ref{62a}) is the order of $\hsig \in \Z_\l$ and (\ref{62c}) says that the sectors of the WZW permutation orbifold $A_g(S_N)/S_N$ are labelled (see also Ref.\,\cite{Halpern:2000vj}) by the ordered partitions $\vec{\s}$ of $N$.

Then the general left-mover current algebra $\hgs$  for the permutation orbifolds is the {\em general semisimple orbifold affine algebra} \cite{deBoer:MBH01,MBHObers}
\begin{subequations}\label{glucks}
\label{csoal}
\begin{eqnarray}
   [ \hj_{\hat{j} aj } (m + \srac{\hat{j}}{\rho (\s)}),
 \hj_{\hat{l} bl} (n + \srac{\hat{l}}{\rho (\s)})]  & = &
 \delta_{jl} \Big\{ if_{ab}{}^c \hj_{\hat{j} + \hat{l},cj }
 (m +n+ \srac{\hat{j} + \hat{l}}{f_j(\s)}) \nn \\
\label{coalg}
 & & \quad\quad 
+  \hat{k}_j(\s) \eta_{ab} (m + \srac{\hat{j}}{\rho (\s)}) \
\delta_{m+n + \frac{\hat{j} + \hat{l}}{f_j(\s)},0}\ \Big\} \;\;\;\;\;\;\;\;\;
\end{eqnarray}
\vspace{-.2in}
\begin{gather}
 \hat{k}_j(\s) \equiv k f_j (\s),  \quad \bar{\hat{j}} =  0 , \ldots , f_j (\s)-1 \, , \qquad \bar{\hat{l}} =  0 , \ldots , f_l (\s)-1 \, \quad   \\
 a,b,c    =  
 1, \ldots, {\rm dim}\,  \g, \quad \s = 0 , \ldots , N_c-1
\end{gather}
\end{subequations}
where $f_{ab}{}^c$ and $\eta_{ab}$ are respectively the structure constants and Killing metric of the untwisted Lie algebra $\gfrak$.
Note that the indices $j,l$ which label the disjoint cycles of $h_\s \in H(\text{permutation})$ are also the semisimplicity indices of the algebra (\ref{csoal}). The quantity $\hat{k}_j(\s)$ is the {\em orbifold affine level} of the $j$th subalgebra of sector $\s$. We also mention the integral affine subalgebra of (\ref{glucks})
\begin{gather}\label{intaffalg}
   [ \hj_{0 aj } (m),
 \hj_{0 bl} (n)]   = 
 \delta_{jl} \Big( if_{ab}{}^c \hj_{0cj }
 (m +n) + \hat{k}_j(\s) \eta_{ab} m  \delta_{m+n,0} \Big),  \quad\s = 0, \dots, N_c-1
\end{gather}
which is an ordinary semisimple affine algebra. Orbifold affine algebras were first seen in Ref.\,\cite{Borisov:1998nc}.

The left-mover twisted affine-Sugawara construction $\hat{T}_\hgs$ of sector $\s$ and its ground state (twist-field state) conformal weight $\hat{\Delta}_0(\s)$
\begin{subequations}\label{gstress}
\begin{gather}
\hat{T}_{\hgs} (z) = \frac{\eta^{ab}}{2k + Q_{\g}} \sum_j \frac{1}{f_j(\s)} \sum_{\hat{j} = 0}^{f_j(\s) -1} : \hj_{\hat{j} a j}(z,\s) \hj_{-\hat{j}, bj}(z,\s) : \, ,\quad a, b = 1, \dots \dim \g\, \\
\hat{c}_{\hgs} = Kc_{\g}, \quad  c_{\g} = \frac{x_\g \dim \g}{x_\g+\tilde{h}_{\g}}, \quad x_\g = \frac{2k}{\psi_\g^2}, \quad \tilde{h}_\g = \frac{Q_\g}{\psi_\g^2}\\
\hat{\Delta}_0(\s) = \frac{c_{\g}}{24}  \sum_j \Big( f_j(\s) - \frac{1}{f_j(\s)}\Big)= \frac{c_{\g}}{24} \Big(K- \sum_j \frac{1}{f_j(\s)}\Big),  \quad\s = 0, \dots, N_c-1
\end{gather}
\end{subequations}
were also given in Refs.\,\cite{deBoer:MBH01, MBHObers}. These references also give the M or mode-ordered form of $\hat{T}_{\hgs}$, which shows the ground state conformal weight more explicitly. These results were first given for the special case $H (\text{permutation})= \Z_\lambda$ (see (\ref{62a})) in Ref.\,\cite{deBoer:1999na}. Here $x_\g$ and $\tilde{h}_\g$ are respectively the invariant level and dual Coxeter number of $\g$, and we note that the stress tensor $\hat{T}_{\hgs}$ is separable into commuting terms $j$.

We turn next to the special features of the action formulation \cite{deBoer:MBH01, MBHObers} of the WZW permutation orbifolds, starting with the explicit form of the twisted representation matrices $\st_{g(\s)} = \{{\cal T}_{\hat{j} a j}\}$ and the orbifold Lie algebra $g(\s)$ in this case
\begin{subequations}\label{formofTperm}
\begin{gather}\label{spec_alg}
\big[\st_{\hat{j} aj}(T, \s), \st_{\hat{l} bl}(T, \s) \big] = \delta_{jl} \, if_{ab}{}^c \st_{\hat{j} + \hat{l}, c j}(T, \s)\, 
\end{gather}
\be\label{factort}
\st_{\hat{j} a j}(T,\s) = T_a t_{\hat{j} j}(\s),  \quad\quad\quad \big[ T_a, T_b \big] = i f_{ab}{}^c T_c, \quad a,b,c = 1, \dots, \dim \g
\e
\begin{gather}\label{t-expl}
 \st_{\hat{j} a j} (T, \s)_{\hat{l} \alpha l}{}^{\hat{m} \beta m} = (T_a)_\alpha{}^\beta t_{\hat{j} j}(\s)_{\hat{l}l}{}^{\hat{m} m }\, 
\, ,\quad \quad t_{\hat{j} j}(\s)_{\hat{l} l}{}^{\hat{m} m} = \delta_{jl} \delta_l{}^m \delta_{\hat{j} + \hat{l} - \hat{m}, 0 \rmod f_j(\s)}\\
 E(T,\s)_{\hat{j} j}{}^{\hat{l} l} = \delta_j{}^l \delta_{\hat{j} - \hat{l}, 0 \rmod f_j(\s)} e^{- 2\pi i \srac{\hat{j}}{f_j(\s)}}, \quad e^{2\pi i \srac{\hat{j}}{f_j(\s)}} t_{\hat{j} j}(\s)_{\hat{l}l}{}^{\hat{m} m} = t_{\hat{j} j}(\s)_{\hat{l}l}{}^{\hat{m} m} e^{-2 \pi i \srac{\hat{l}-\hat{m}}{f_l(\s)}} \label{selr1}
\end{gather}
\begin{gather}
\bar{\hat{j}}, \bar{\hat{l}}, \bar{\hat{m}} = 0, \dots, f_j(\s)-1, \quad \alpha, \beta = 1, \dots, \dim T, \quad \s = 0, \dots, N_c-1
\end{gather}
\end{subequations}
where $T_a$ is any irrep of $\g$ and the last relation in (\ref{selr1}) is the form taken by the $\st$-selection rule (\ref{sel-rule-for-st-1}) in the case of the WZW permutation orbifolds.
As noted more generally above, the structure constants ${\cal F}_{\hgs}$ of ${g}(\s)$ in (\ref{spec_alg}) are the same as those of the twisted affine algebra $\hgs$ in (\ref{csoal}).

For the WZW permutation orbifolds the factorized form $\st = Tt(\s)$ of the twisted representation matrices in (\ref{formofTperm}) shows that the group orbifold elements $\hat{g}$ are direct sums \cite{deBoer:MBH01}
\begin{subequations} \label{gfact}
\begin{gather}
\hat{g}(\st, \xi, t, \s) = e^{i \hat{\beta}^{\hat{j} aj}(\xi, t, \s) \st_{\hat{j} a j} (T,\s)} \,\label{g_ex_perm} \\
\hat{g}(\st, \xi, t, \s)_{\hat{j} \alpha j}{}^{\hat{j}' \beta j'} = \delta_j{}^{j'} \hat{g}_j(\st, \xi, t, \s)_{\hat{j} \alpha }{}^{\hat{j}' \beta }\,,  \qquad 
\bar{\hat{j}}, \bar{\hat{j}}' = 0, \dots, f_j(\s)-1 \label{gblocks}\\
\hat{g}_j(\st, \xi, t, \s) \equiv  e^{i\sum_{\hat{j}, a} \hat{\beta}^{\hat{j} a j}(\xi, t,\s) T_a \tau_{\hat{j}}(j,\s)}, \label{gfact2}\\  
\tau_{\hat{j}}(j,\s)_{\hat{j}'}{}^{\hat{j}''} \equiv \delta_{\hat{{j}} + \hat{{j}}' - \hat{{j}}'', 0 \rmod f_j(\s)} \,, \; \bar{\hat{j}},\bar{\hat{j}}',\bar{\hat{j}}''\!\!\!= 0, \dots, f_j(\s)-1\label{taudefi}\\
  e^{2\pi i \srac{\hat{j}}{f_j(\s)}} \tau_{\hat{j}}(j,\s)_{\hat{j}'}{}^{\hat{j}''} = \tau_{\hat{j}}(j,\s)_{\hat{j}'}{}^{\hat{j}''} e^{-2 \pi i \srac{\hat{j}'-\hat{j}''}{f_l(\s)}}\label{tausel}
\end{gather}
\end{subequations}
where $\{\hat{\beta}^{\hat{j} a j}\}$ are the twisted tangent space coordinates in this case. The block structure in (\ref{gblocks}) reflects the semisimplicity of the twisted current algebra $\hgs$ in (\ref{glucks}).
Then the monodromies of these objects
\begin{subequations}\label{gmonoperm}
\begin{gather}
\hat{\beta}^{\hat{j} a j}(\xi + 2 \pi, t, \s) = \hat{\beta}^{\hat{j} a j}(\xi, t, \s) e^{2\pi i \frac{\hat{j}}{f_j(\s)}}  \\
\hat{g}(\st, \xi + 2\pi, t, \s)_{\hat{j} \alpha j}{}^{\hat{l} \beta l} = e^{-2\pi i \frac{\hat{j} - \hat{l}}{f_j(\s)}} \hat{g}(\st, \xi , t, \s)_{\hat{j} \alpha j}{}^{\hat{l} \beta l}= e^{-2\pi i \frac{\hat{j} - \hat{l}}{f_l(\s)}} \hat{g}(\st, \xi , t, \s)_{\hat{j} \alpha j}{}^{\hat{l} \beta l}\\
\hat{g}_j(\st, \xi + 2\pi, t, \s)_{\hat{j} \alpha }{}^{\hat{j}' \beta } = e^{-2\pi i \frac{\hat{j} - \hat{j}'}{f_j(\s)}} \hat{g}_j(\st, \xi , t, \s)_{\hat{j} \alpha }{}^{\hat{j}' \beta } \label{nochmehr}
\end{gather}
\end{subequations}
follow from Eqs. (\ref{gmon}), (\ref{b-mono}) and (\ref{permlabel}). The consistency of these relations follows from the forms of the $\st$-selection rule in (\ref{selr1}), (\ref{tausel}) and the fact that the group orbifold element $\hat{g}$ is a direct sum.

To obtain a more explicit form of the WZW orbifold action (\ref{gorbact}) for the permutation orbifolds,  we use the direct sum form of the group orbifold element $\hat{g}$ and the explicit form of the twisted data matrix \cite{deBoer:MBH01} 
\be
{\cal M} (\st ( T,\s), \s)_{\hat{j} \alpha j}{}^{\hat{j}' \beta j'} = \frac{k}{y(T)} \delta_\alpha{}^\beta \delta_j{}^{j'} \delta_{\hat{j}-\hat{j}', 0 \rmod f_j(\s)}, \quad \bar{\hat{j}}, \bar{\hat{j}}' = 0, \dots, f_j(\s)-1
\e
in sector $\s$ of each WZW permutation orbifold. Then one finds for the WZW permutation orbifolds that
the WZW orbifold action (\ref{gorbact}) is separable in the semisimplicity index $j$ \cite{deBoer:MBH01}:
\begin{subequations}\label{713}
\be
\widehat{S}_{\hg(\s)}[\M,\hg] = \sum_j \widehat{S}_{\s,j}[\hg_j], \quad \hg_j \equiv \hg_j(\st(T,\s), \xi, t,\s), \quad \forall\,T\quad \text{for each }\s = 0, \dots, N_c-1
\e
\vspace{-0.2in}
\begin{gather}\label{S_j}
\widehat{S}_{\s,j}[\hg_j] \equiv - \frac{k}{y(T)} \sum_{\hat{j} = 0}^{f_j(\s) -1}\sum_{\alpha=1}^{\dim T} [\frac{1}{8\pi} \int d^2\xi \; \hat{g}_j^{-1} \partial_+ \hat{g}_j \hat{g}^{-1}_j \partial_- \hg_j +\frac{1}{12 \pi} \int_{\Gamma}
(\hat{g}_j^{-1} d\hat{g}_j)^3]_{\hat{j} \alpha}{}^{\hat{j} \alpha} \\
\widehat{S}_{\s,j}[\hat{g}_j(\st, \xi + 2\pi,t,\s)] = \widehat{S}_{\s,j}[\hat{g}_j(\st, \xi,t,\s)], \; \forall \, j .
\end{gather}
\end{subequations}
This separability describes an independent dynamics for each block $\hat{g}_j$, which corresponds to  the semisimplicity of the general orbifold affine algebra (\ref{csoal}) and the separability of the twisted affine-Sugawara constructions in (\ref{gstress}).

\section{Example: The Permutation Coset Orbifolds}\label{sect:permcoset}

We are now prepared to consider the {\em general permutation coset orbifold}
\begin{subequations}
\begin{gather}
\frac{A_{g/h}(H)}{H}, \quad H(\text{permutation}) \subset S_N \\
g = \bigoplus_{I=0}^{K-1} \g^I, \quad \g^I \cong \g, \quad T^I_a \cong T_a, \quad k_I = k \\
H(\text{permutation}) \subset Aut(g), \quad H(\text{permutation}) \subset Aut(h)
\end{gather}
\end{subequations}
where $H(\text{permutation})$ acts among the copies of $\g$, and $h = h(H)$ is an $H(\text{permutation})$-covariant subalgebra of $g$. The stress tensors $\hat{T}_{\hgs/\hhs}$ of the permutation coset orbifolds have the form  $\hat{T}_{\hgs} - \hat{T}_\hhs$ given in Eq. (\ref{stressT_gauge}), where the orbifold affine algebra $\hgs$ and the stress tensor $\hat{T}_\hgs$ are given in Eqs. (\ref{glucks}) and (\ref{gstress}) respectively, and $\hhs \subset \hgs$ is the appropriate twisted affine subalgebra of $\hgs$. 

This set of orbifolds includes the relatively simple case of the {\em permutation coset copy orbifolds} (see Sec.\,\ref{sect:copy}) and
 the more involved cases of the {\em interacting permutation coset orbifolds} (see Sec.\,\ref{sect:diag}). The hallmark of the coset copy orbifolds is that their stress tensors are separable into commuting terms $j$
\be\label{hllm}
\hat{T}_{\hgs/\hhs} = \sum_j  \hat{T}_{\hgs/\hhs}^j
\e 
while the stress tensors of the interacting coset orbifolds show interaction among the different values of $j$. Both of these cases were discussed in some detail for $H(\text{permutation}) = \Z_{\l}$ in Ref.\,\cite{Evslin:1999ve}. In this section we discuss the form taken by the general coset orbifold action (\ref{empty_+}) in the special case of the permutation coset orbifolds.

For WZW permutation orbifolds, the monodromy of the group orbifold elements $\hat{g}$ was given in Eq. (\ref{gmonoperm}). As discussed more generally in Sec.\,\ref{sect:cosetaction}, the monodromies of the subgroup orbifold elements $\hat{h}$ and their associated quantities have the same form as that of the group orbifold element $\hat{g}$
\begin{subequations}\label{hmonop}
\begin{gather}\label{hmono2}
\hat{O}(\st, \xi+2\pi, t,\s) = E(T,\s) \hat{O}(\st, \xi, t,\s) E(T,\s)^*, \quad \hat{O} = \hat{g}, \hat{h}_\pm, \hat{A}_{\pm}, \text{ or } \hat{\psi}\\
\hat{O}(\st, \xi+2\pi, t,\s)_{\hat{j} \alpha j}{}^{\hat{j}' \beta j'} = e^{-2 \pi i \frac{ \hat{j} -\hat{j}'}{f_j(\s)}} \hat{O}(\st, \xi, t, \s)_{\hat{j} \alpha j}{}^{\hat{j}' \beta j'}\\
\bar{\hat{j}}, \bar{\hat{j}}' = 0, \dots, f_j(\s)-1 ,  \quad\s = 0, \dots, N_c-1
\end{gather}
\end{subequations}
although the monodromies of $\hat{h}, \hat{A}_\pm$ and $\hat{\psi}$ can look quite different when the support of $\hat{h}$ is taken into account (see the examples in Secs.\,\ref{sect:copy}, \ref{sect:diag} and \ref{sect:other_ex}).
Moreover, the subgroup orbifold elements $\hat{h}\subset \hat{g}$ and their associated quantities are block diagonal
\begin{subequations}\label{blocks}
\begin{gather}
\hat{h}(\st, \xi, t, \s)_{\hat{j} \alpha j}{}^{\hat{j}' \beta j'} = \delta_j{}^{j'} \hat{h}_j(\st, \xi, t, \s)_{\hat{j} \alpha }{}^{\hat{j}' \beta }\, \in \widehat{\HH}(\s)\label{antriv0} \\ 
\hat{A}_\pm(\st, \xi, t, \s)_{\hat{j} \alpha j}{}^{\hat{j}' \beta j'} =  \delta_j{}^{j'} \hat{A}_{\pm,j}(\st, \xi, t, \s)_{\hat{j} \alpha }{}^{\hat{j}' \beta } \, \in \hat{h}(\s)\\
\hat{A}_{+,j} (\st\!, \xi, t, \s)=\! - i \hat{h}_{+,j}^{-1}(\st\!, \xi, t, \s)\partial_+ \hat{h}_{+,j}(\st\!, \xi, t, \s)\, \label{blocks_1}\\
 \hat{A}_{-,j} (\st\!, \xi, t, \s) =\! - i  \hat{h}_{-,j}(\st\!, \xi, t, \s) \partial_-\hat{h}_{-,j}^{-1}(\st\!, \xi, t, \s)\label{blocks_2}\\
\hat{\psi}(\st, \xi, t, \s)_{\hat{j} \alpha j}{}^{\hat{j}' \beta j'} =  \delta_j{}^{j'} \hat{\psi}_j(\st, \xi, t, \s)_{\hat{j} \alpha }{}^{\hat{j}' \beta } \, \in \widehat{\HH}(\s)
\end{gather}
\end{subequations}
because, for WZW permutation orbifolds, all group orbifold elements $\hat{g}$ are block diagonal (see Eq.\,(\ref{gfact})). Then we find the monodromies of the $j$ blocks
\begin{gather}
\hat{O}_j(\st, \xi + 2\pi, t, \s)_{\hat{j} \alpha}{}^{\hat{j}' \beta} = e^{-2\pi i \frac{\hat{j} - \hat{j}'}{f_j(\s)}} \hat{O}_j(\st, \xi , t, \s)_{\hat{j} \alpha}{}^{\hat{j}' \beta}, \quad \hat{O}_j = \hat{g}_j, \hat{h}_{\pm,j},  \hat{A}_{\pm, j} \text{ or } \hat{\psi}_j 
\end{gather}
which include the monodromy (\ref{nochmehr}) of the blocks $\hat{g}_j$. The twisted gauge transformation of the blocks
\begin{subequations}
\begin{gather}
\hat{g}_j(\st, \xi, t, \s)^{\hat{\psi}_j} = \hat{\psi}_j (\st,\xi, t, \s) \hat{g}_j (\st,\xi, t, \s) \hat{\psi}_j^{-1}(\st,\xi, t, \s)\; \\
\hat{h}_{+,j}(\st, \xi, t, \s)^{\hat{\psi}_j} = \hat{h}_{+,j}(\st, \xi, t, \s) \hat{\psi}_j^{-1}(\st, \xi, t, \s), \quad  \hat{h}_{-,j}(\st, \xi, t, \s)^{\hat{\psi}_j} =  \hat{\psi}_j(\st, \xi, t, \s) \hat{h}_{-,j}(\st, \xi, t, \s) \\
\hat{A}_{\pm,j}(\st\!\!,\xi, t, \s)^{\hat{\psi}_j}  = \hat{\psi}_j(\st\!\!,\xi, t, \s) \hat{A}_{\pm, j}(\st\!\!,\xi, t, \s) \hat{\psi}_j^{-1}(\st\!\!,\xi, t, \s) + i \partial_\pm \hat{\psi}_j(\st\!\!,\xi, t, \s) \hat{\psi}^{-1}_j(\st\!\!,\xi, t, \s)
\end{gather}
\end{subequations}
also follow from Eqs. (\ref{twgauge_trafo}) and (\ref{blocks}).

Then the general coset orbifold action (\ref{empty_+}) reduces in the case of the general permutation coset orbifold to the following separable form  
\begin{subequations}\label{moreactions}
\begin{gather}
\widehat{S}_{\hg(\s)/\hat{h}(\s)}[\M, \hg, \hat{A}_{\pm}] =  \sum_{j}  \Big( \widehat{S}_{\s,j} [ \hat{g}_j] + \widehat{S}_{\s,j}[\hg_j, \hat{A}_{\pm,j}]\Big)\\
 \hat{g}_j\equiv \hat{g}_j(\st(T,\s), \xi, t,\s), \quad \hat{A}_{\pm,j} \equiv \hat{A}_{\pm,j} (\st(T,\s), \xi, t,\s),\quad \forall\,T\quad \text{for each }\s = 0, \dots, N_c-1 
\end{gather}
\begin{gather}
 \widehat{S}_{\s,j}[\hg_j, \hat{A}_{\pm,j}] \equiv 
-\frac{k}{4\pi y(T)} \sum_{\hat{j} = 0}^{f_j(\s)-1} \sum_{\alpha=1}^{\dim T} \int d^2\xi\, [
i \hg_j^{-1} \partial_+ \hg_j \hat{A}_{-,j} -  i \hat{A}_{+,j} \partial_- \hat{g}_j \hat{g}^{-1}_j \quad\quad\quad\quad\quad\quad \nn  \\
\vspace{-0.6in}
 \quad\quad\quad\quad\quad\quad\quad\quad\quad\quad\quad\quad\quad\quad\quad\quad\quad\quad - \hg_j^{-1}\hat{A}_{+,j} \hg_j \hat{A}_{-,j} + \hat{A}_{+,j} \hat{A}_{-,j}]_{\hat{j}\alpha}{}^{\hat{j} \alpha} 
\end{gather}
\begin{gather}
\widehat{S}_{\s,j}[\hat{g}_j(\st, \xi + 2\pi,t,\s), \hat{A}_{\pm, j}(\st, \xi + 2\pi,t,\s)] = \widehat{S}_{\s,j}[\hat{g}_j(\st, \xi,t,\s), \hat{A}_{\pm, j}(\st, \xi,t,\s)]\\  
\widehat{S}_{\s,j}[\hat{g}_j^{\hat{\psi}_j}, \hat{A}_{\pm,j}^{\hat{\psi}_j}] = 
\widehat{S}_{\s,j}[\hat{g}_j, \hat{A}_{\pm,j}]
\end{gather}
\end{subequations}
where $\widehat{S}_{\s,j} [ \hat{g}_j]$ is defined in Eq. (\ref{713}).  In this case however separability is not necessarily the same as independent dynamics for each $j$ block --- because the twisted gauge field blocks $\{\hat{A}_{\pm, j}\}$ are not generically independent quantities. In special cases however (see e.g. Sec.\,\ref{sect:copy}) the twisted gauge field blocks may be independent and independent dynamics is obtained for each block $\hat{g}_j$.

We finally remark on the tangent space description of the permutation coset orbifolds.  Using the explicit form of the twisted representation matrices $\st(T,\s) =  T t(\s)$ in (\ref{formofTperm}), the tangent space description of  the group orbifold elements $\hg$ of the WZW permutation orbifolds was given in (\ref{gfact}). The recipe of Sec.\,\ref{sect:twistedsubal} then gives the tangent space description of the subgroup orbifold elements $\hat{h}$
\begin{subequations}\label{antriv}
\begin{gather}
\hat{h}(\st(T,\s), \xi, t, \s) = e^{i \hat{\beta}_{\hat{h}(\s)}(\xi, t, \s) \cdot \st_{h(\s)}(T,\s)} =  
e^{i\{ \hat{\beta}_{\hat{h}(\s)}(\xi, t, \s)\cdot{\cal E}(\s) \}\cdot T t(\s)}\in \widehat{\HH}(\s) \subset \widehat{\G}(\s)\\ 
\forall \hat{\beta}_{\hat{h}(\s)}, \;\forall\, T \quad 
\text{for each } \srange
\end{gather}
\end{subequations}
where the embedding matrix ${\cal E}(\s)$ remains to be determined in each subexample. We find it interesting that the form (\ref{antriv}) must always be consistent with the block-diagonal form of $\hat{h}$ in (\ref{antriv0}). 

This completes our discussion of the general permutation coset orbifold, and we turn now to work out some simple sub-examples of this type in further detail.

\section{Sub-Example: The Permutation Coset Copy Orbifolds}\label{sect:copy}

The case of {\em cyclic} coset copy orbifolds $
{\bigoplus_{I=0}^{\lambda -1} \Big(\frac{\mathfrak{g}}{\mathfrak{h}}\Big)^I}/{\Z_\lambda}$
was considered at the current-algebraic level in Refs.\,\cite{Borisov:1998nc} and \cite{Evslin:1999ve}. Here we extend the results of Ref.\,\cite{Evslin:1999ve} to include both the current-algebraic formulation and the action formulation of {\em all} permutation coset copy orbifolds. 

We begin with a set of $K$ commuting copies of a coset construction $\gfrak/\hfrak$ where $\G/\HH$ is a reductive coset space
\begin{subequations}
\begin{gather}
g = \bigoplus_{I=0}^{K-1} \gfrak^I, \quad h = \bigoplus_{I=0}^{K-1} \hfrak^I, \quad 
\gfrak^I \cong \gfrak, \quad \hfrak^I \cong \hfrak, \quad \hfrak \subset \gfrak, \quad K \le N \\
 \frac{g}{h} = \bigoplus_{I=0}^{K-1} \Big(\frac{\g}{\h}\Big)^I, \quad \Big(\frac{\g}{\h}\Big)^I\cong \frac{\g}{\h}\label{dirsum1} .
\end{gather}
\end{subequations}
The symmetry group $H(\text{permutation})\subset S_N$ permutes the copies $\gfrak^I$ and $\hfrak^I$ of $\gfrak$ and $\hfrak$, so that 
\be
H(\text{permutation}) \subset Aut(g), \quad H(\text{permutation}) \subset Aut(h) .
\e
This means in particular that $h = h(H)$ is an {\em $H(\text{permutation})$-covariant subalgebra} of $g$ 
and the coset construction  $g/h$ in (\ref{dirsum1}) is an $H(\text{permutation})$-invariant coset construction. For these coset constructions, the untwisted embedding $h \subset g$ has the simple form
\begin{gather}\label{affEmb}
J_{AI}(z) = {\cal E}(0)_A{}^a J_{aI}(z), \quad 
I= 0, \dots, K-1, \quad a = 1, \dots, \dim \gfrak, \quad A \in \hfrak
\end{gather}
where the affine embedding matrix ${\cal E}(0)$ is $I$-independent. 

Then we may consider the {\em general permutation coset copy orbifold}
\be
\frac{\frac{g}{h}}{H(\text{permutation})}
= \frac{\bigoplus^{K-1}_{I=0}(\frac{\g}{\h})^I}{H(\text{permutation})} \, 
\e
as a special case of the permutation coset orbifolds.
Defining eigencurrents ${\cal J}$, it follows \cite{Evslin:1999ve} by local isomorphisms ${\cal J} \dual \hj$ that
\begin{subequations}\label{aff_emb}
\begin{gather}
\hj_{\hat{j}Aj}(z,\s) = {\cal E}(0)_A{}^a \hj_{\hat{j} a j}(z,\s)\label{aff_emba} \\
\hj_{\hat{j}aj}(z e^{2\pi i}, \s)  = e^{-2 \pi i \frac{ \hat{j}}{f_j(\s)}} \hj_{\hat{j}aj}(z, \s), \qquad 
\hj_{\hat{j}Aj}(z e^{2\pi i}, \s)  = e^{-2\pi i \frac{ \hat{j}}{f_j(\s)}} \hj_{\hat{j}Aj}(z,\s)\\
a = 1, \dots, \dim \gfrak, \quad A \in \hfrak, \quad \bar{\hat{j}} = 0, \dots, f_j(\s) -1, \quad \s = 0, \dots, N_c-1
\end{gather}
\end{subequations}
where $\hj_\hgs = \{\hj_{\hat{j} a j}\}$ are the twisted currents of the ambient twisted affine algebra $\hgs$ and the twisted affine embedding matrix has the form ${\cal E}(\s) = {\cal E}(0)$ in this case.
The twisted currents $\hj_{\hhs} = \{\hj_{\hat{j} A j}\}$ satisfy the twisted affine subalgebra  $\hhs$ in (\ref{affine_sub}) and, in the special case when ${\cal E}(0)_a{}^A = \delta_a{}^A$ the twisted affine subalgebra $\hhs$ is just Eq. (\ref{coalg}) with $a,b,c \rightarrow A,B,C$. 

Following Refs.\,\cite{Evslin:1999ve,Halpern:2000vj, deBoer:MBH01,MBHObers} we find the left-mover stress tensor and ground state (twist-field state) conformal weight
\begin{subequations}\label{stress_exa1}
\begin{gather}\label{morestress1}
\hat{T}_{\srac{\hat{\mathfrak{g}}(\s)}{\hat{\mathfrak{h}}(\s)}} (z) = \hat{T}_{\hat{\mathfrak{g}}(\s)} (z) - \hat{T}_{\hat{\mathfrak{h}}(\s)} (z)  = \sum_j \sum_{\hat{j} = 0}^{f_j(\s)-1}{\cal L}_{\frac{\hgs}{\hhs}}^{\hat{j} a j; -\hat{j}, b j} (\s) : \hj_{\hat{j} a j}(z, \s) \hj_{-\hat{j}, bj}(z, \s) :
\end{gather}
\begin{gather}\label{morestress2}
\hat{T}_{\hgs} (z) = \frac{\eta^{ab}}{2k + Q_{\g}} \sum_j \frac{1}{f_j(\s)} \sum_{\hat{j} = 0}^{f_j(\s) -1} : \hj_{\hat{j} a j}(z, \s) \hj_{-\hat{j}, bj}(z, \s) : \, ,\quad a, b = 1, \dots \dim \g\, \\
\hat{T}_{\hhs} (z) = \frac{\eta^{AB}}{2k + Q_{\h}} \sum_j\frac{1}{f_j(\s)}\sum_{\hat{j} = 0}^{f_j(\s) -1} : \hj_{\hat{j} A j}(z, \s) \hj_{-\hat{j}, Bj}(z, \s) : \, ,\quad A,B \in \hfrak \label{morestress3}\\
\hat{c}_{\hgs/\hhs}=c_{g/h}
 = \lambda x_{\g}(\frac{\dim \gfrak}{x_{\g}+\tilde{h}_{\h}}
-\frac{r\dim\hfrak}{rx_{\g}+\tilde{h}_{\h}}), \qquad \hat{\Delta}_0(\s)=\frac{c_{g/h}}{24}(K-\sum_j \frac{1}{f_j(\s)}) \\
\hat{T}_{\frac{\hgs}{\hhs}}(z) \hj_{\hat{j}Aj}(w, \s) = {\cal O}((z-w)^0), \quad \forall \; A \in \mathfrak{h},  \quad\s = 0, \dots, N_c-1
\end{gather}
\end{subequations}
for each sector $\s$ of each permutation coset copy orbifold. This result was first given for the special case $H(\text{permutation}) = \Z_\l$ (see (\ref{62a})) in Ref.\,\cite{Evslin:1999ve}. Here $\eta^{AB}$ is the induced inverse Killing metric of the untwisted Lie algebra $\hfrak \subset \gfrak$ and $r$ is the index of embedding of $\hfrak$ in $\gfrak$. The explicit form of the twisted inverse inertia tensor ${\cal L}_{\hgs/\hhs}(\s)$ in (\ref{morestress1}) is easily worked out from the forms of $\hat{T}_\hgs$ and $\hat{T}_\hhs$ in (\ref{morestress2}) and (\ref{morestress3}). 
Note that for these simple coset orbifolds the stress tensor $\hat{T}_{\hgs/\hhs}$ is separable into commuting and therefore non-interacting terms $j$. This is in contrast to the forms of  $\hat{T}_{\hgs/\hhs}$ found for the interacting permutation coset orbifolds (see Sec.\,\ref{sect:diag}).



Following the recipe of Sec.\,\ref{sect:twistedsubal}, the affine embedding (\ref{aff_emba}) also  gives us the twisted representation matrices ${\cal T}_{h(\s)} = \{\st_{\hat{j} Aj}(T,\s)\}$ of the orbifold Lie subalgebra $h(\s)$ in this case
\begin{gather}\label{formofTcopy}
\st_{\hat{j} A j}(T,\s) = {\cal E}(0)_A{}^a \st_{\hat{j} a j}(T,\s) = T_A t_{\hat{j}j}(\s),  \quad T_A = {\cal E}(0)_A{}^a T_a, \quad 
 a=1, \dots, \dim \g, \quad A \in  \hfrak \subset \gfrak\,   
\end{gather}
where we have used the factorized form of ${\cal T} $ in (\ref{formofTperm}) and recognized the Lie embedding $T_A$, $A \in \hfrak$ implied by the affine embedding (\ref{affEmb}). The general subgroup orbifold element $\hat{h} \in \widehat{\HH}(\s) \subset \widehat{\G}(\s)$ in (\ref{antriv}) then takes the form
\begin{subequations}\label{hcp}
\begin{align}
\hat{h}(\st, \xi, t, \s)  &=  
e^{i \hat{\beta}_{\hat{h}(\s)}(\xi, t, \s)\cdot \st_{h(\s)} (T,\s)}  =  e^{i\sum_{\hat{j} a j} \big\{\sum_A \hat{\beta}^{\hat{j} Aj}(\xi, t, \s) {\cal E}(0)_A{}^a\big\}T_a t_{\hat{j}j}(\s)}\\
 &= e^{i \sum_{\hat{j} Aj}\hat{\beta}^{\hat{j} Aj}(\xi, t, \s) T_A t_{\hat{j} j} (\s)}, \quad \forall \hat{\beta}^{\hat{j} A j} ,  \quad\forall T \; \text{for each }\s = 0, \dots, N_c-1 \label{hexa1-1}
\end{align}
\end{subequations}
where $\{ \hat{\beta}^{\hat{j} A j}\}$ are the tangent space coordinates on $\hat{h}(\s)$ and $\{\hat{\beta}^{\hat{j} Aj}(\xi, t, \s) {\cal E}(0)_A{}^a\}$ is the restriction to $\hat{h}(\s)$ of the tangent space coordinates $\hat{\beta}^{\hat{j} aj}$ on $\hat{g}(\s)$.
The monodromy of $\hat{\beta}^{\hat{j} Aj}$ is 
\begin{gather}
\hat{\beta}^{\hat{j} A j}(\xi + 2 \pi, t, \s) =  \hat{\beta}^{\hat{j} A j} (\xi, t,\s)e^{ 2\pi i \frac{\hat{j}}{f_j(\s)}}, \quad A \in \hfrak,  \quad\s = 0, \dots, N_c-1\label{beta_mono}
\end{gather}
and it is not difficult to verify with (\ref{t-expl}) that this monodromy is equivalent to the monodromy (\ref{hmono2}) of the subgroup orbifold element $\hat{h}$. 

Using the form of $\hat{h}$ in  (\ref{hcp}) and the block decompositions (\ref{gfact2}) and (\ref{blocks}), we find the forms of the blocks for the permutation coset copy orbifolds 
\begin{subequations}\label{haps_fact}
\begin{gather}
\hat{g}_j(\st, \xi, t, \s) = e^{i\sum_{\hat{j}, a} \hat{\beta}^{\hat{j} a j}(\xi, t, \s) T_a \tau_{\hat{j}}(j,\s)}\, , \quad \hat{h}_{\pm, j}(\st, \xi, t, \s) = e^{i\sum_{\hat{j}, A} \hat{\beta}_\pm^{\hat{j} A j}(\xi, t, \s) T_A \tau_{\hat{j}}(j,\s)}\label{a_fact}\\
\hat{A}_{\pm , j}(\st, \xi, t,\s)_{\hat{j}' \alpha}{}^{\hat{j}'' \beta} = \sum_A \sum_{\hat{j} = 0}^{f_j(\s)-1} \hat{A}_{\pm, j}{}^{A \hat{j}}(\xi, t, \s) \big( T_A\big)_\alpha{}^\beta \tau_{\hat{j}}(j,\s)_{\hat{j}'}{}^{\hat{j}''} \label{A_blah}\\
 \hat{\psi}_j(\st, \xi, t, \s) = e^{i \sum_{\hat{j}, A}\hat{\gamma}^{\hat{j} A j}(\xi, t,\s) T_A \tau_{\hat{j}}(j,\s)}
\end{gather}
\end{subequations}
where the matrices $\tau_{\hat{j}}(j,\s)$ are defined in (\ref{taudefi}). The coefficient $\hat{A}_{\pm, j}{}^{A\hat{j}}$ in (\ref{A_blah}) is determined by (\ref{hexa1-1}), (\ref{blocks_1}) and (\ref{blocks_2}).
We remark that each twisted gauge field block $\hat{A}_{\pm, j}$ is {\em independent} in this case because $\hat{\beta}_\pm^{\hat{j} A j}$ can be specified independently for each $j$. 

The coset orbifold action for the permutation coset copy orbifolds has the form given in (\ref{moreactions}) 
\begin{subequations}
\begin{gather}
\widehat{S}_{\hg(\s)/\hat{h}(\s)}[\M, \hg, \hat{A}_{\pm}] = \sum_{j}  \Big( \widehat{S}_{\s,j} [ \hat{g}_j] + \widehat{S}_{\s,j}[\hg_j, \hat{A}_{\pm,j}]\Big) \\
\hat{g}_j\equiv \hat{g}_j(\st(T,\s), \xi, t,\s), \quad \hat{A}_{\pm,j} \equiv \hat{A}_{\pm,j} (\st(T,\s), \xi, t,\s), \quad \forall\, T\quad \text{for each } \s = 0, \dots, N_c-1\\
\widehat{S}_{\s,j}[\hat{g}_j(\st, \xi + 2\pi,t,\s), \hat{A}_{\pm, j}(\st, \xi + 2\pi,t,\s)] = \widehat{S}_{\s,j} [\hat{g}_j(\st, \xi,t,\s), \hat{A}_{\pm, j}(\st, \xi,t,\s)] \\
\widehat{S}_{\s,j}[\hat{g}_j^{\hat{\psi}_j}, \hat{A}_{\pm,j}^{\hat{\psi}_j}] = 
\widehat{S}_{\s,j}[\hat{g}_j, \hat{A}_{\pm,j}]  
\end{gather}
\end{subequations}
where the special form of the $\hat{h}$ and $\hat{A}_{\pm}$ blocks for this case are given in (\ref{haps_fact}). For the permutation coset copy orbifolds, integration of the independent gauge fields $\hat{A}_{\pm, j}$ will result in a set of twisted sigma model actions, each having the form $\sum_js_{\s,j}(\hat{g}_j)$  with {\em no coupling between different blocks $\hat{g}_j$}. This decoupling reflects the non-interacting  form (\ref{stress_exa1}) of the stress tensors of these coset orbifolds.


\section{Sub-Example: The Diagonal Permutation Coset Orbifolds}\label{sect:diag}
As another special case of the permutation coset orbifolds, we consider here the simplest set  of {\em interacting}  permutation coset orbifolds, namely the {\em diagonal permutation coset orbifolds}
\begin{subequations}\label{Kdef}
\begin{gather}
\frac{\frac{g}{h}}{H(\text{permutation})} = \frac{\frac{\bigoplus_{I=0}^{K-1} \g^I}{\g_{\text{diag}}}}{H(\text{permutation})}\, \quad \gfrak^I \cong \gfrak \\ H(\text{permutation}) \subset S_N, \quad  H(\text{permutation}) \subset Aut(g), \quad  H(\text{permutation}) \subset Aut(h)
\end{gather}
\end{subequations}
where $K \le N$ is the number of copies of $\g$ in $g$, and $h = \gfrak_{\text{diag}} \subset g$, $\gfrak_{\text{diag}}\cong \gfrak$ is the diagonal subalgebra of $g$. 
The special case of $H(\text{permutation})=\Z_\lambda, K=\l$ was discussed at the twisted current-algebraic level in Refs.\,\cite{Borisov:1998nc,deBoer:1999na, Evslin:1999ve}.

For the diagonal coset constructions, the diagonal subalgebra $\g_{\text{diag}}$ is a very simple type of $H$-covariant subalgebra, namely an {\em $H$-invariant subalgebra}, and the diagonal currents 
\be\label{Jdiag}
J_a^{\text{diag}}(z) \equiv \sum_{I=0}^{K-1} J_{aI}(z)\, , \quad a = 1, \dots, \dim \g
\e
generate affine $h$ $\cong$ affine $\gfrak$ at level $Kk$. The relation (\ref{Jdiag}) provides an implicit definition of the untwisted affine embedding matrix ${\cal E}(0)$. Using an identity in Ref.\,\cite{MBHObers}, we can express the diagonal currents in terms of the eigencurrents ${\cal J}_{\hat{j}aj}(z, \s)$: 
\be
J_a^{\text{diag}}(z) = \sum_{I=0}^{K-1} J_{aI}(z) = \sum_{j} {\cal J}_{0aj}(z, \s),  \quad\s = 0, \dots, N_c-1 .
\e
Then the principle of local isomorphisms ${\cal J} \dual \hj$ gives the corresponding orbifold currents $\hj_\hhs = \{\hj^{\text{diag}}(z, \s)\}$ of $\hhs \subset \hgs$ 
\begin{subequations}\label{firststar}
\begin{gather}
J_a^{\text{diag}}(z) =  \sum_{j} {\cal J}_{0aj}(z,\s) \dual \hat J_a^{\text{diag}}(z,\s) =  \sum_{j}  \hat{J}_{0aj}(z,\s)\label{fs1}\\
\hj_a^{\text{diag}}(ze^{2\pi i },\s) =  \hj_a^{\text{diag}}(z,\s), \quad a = 1, \dots, \dim \g,  \quad\s = 0, \dots, N_c-1
\end{gather}
\end{subequations}
where the modes of $\hat{J}_{0aj}(z,\s)$ are the generators of the integral affine subalgebra (\ref{intaffalg}) of the general orbifold affine algebra (\ref{csoal}). The last relation in (\ref{fs1}) provides an implicit definition of the affine embedding matrix ${\cal E}(\s)$ for the twisted affine embedding $\hhs \subset \hgs$.
The currents $\hj^{\text{diag}}(z,\s)$ of $\hhs$ have trivial monodromy in this case because $h = \gfrak_{\text{diag}}$ is an invariant subalgebra under $H$.
Using (\ref{intaffalg}), we find the explicit form of the twisted affine subalgebra $\hhs$ 
\begin{subequations}
\begin{gather}\label{secondstar}
 \hat J_a^{\text{diag}}(m) =  \sum_{j}  \hat{J}_{0aj}(m), \quad m \in \Z\\
\big[ \hj_a^{\text{diag}}(m), \hj_b^{\text{diag}}(n)\big] = i f_{ab}{}^c \hj_c^{\text{diag}}(m+n) + Kk \eta_{ab} \, m \delta_{m+n, 0}\\  a,b,c = 1, \dots, \dim \gfrak,  \quad\s = 0, \dots, N_c-1
\end{gather}
\end{subequations}
which is another affine $\gfrak$ at level $Kk$. 

It follows that the left-mover stress tensor and ground state (twist-field state) conformal weight of sector $\s$ is\f{The first example of this type of twisted affine-Sugawara construction was given by Kac and Wakimoto \cite{kacwak}. Their example was later identified \cite{Borisov:1998nc,deBoer:1999na} as the stress tensor in each twisted sector $\s = 1, \dots, \lambda-1$ of the interacting cyclic coset orbifolds $(A_g(\Z_\lambda)/g_{\text{diag}})/\Z_\lambda$ for $\lambda$ prime.}
\begin{subequations}\label{Texa1}
\begin{gather}
\hat{T}_{\srac{\hat{\mathfrak{g}}(\s)}{\hat{\mathfrak{h}}(\s)}} (z) = \hat{T}_{\hat{\mathfrak{g}}(\s)} (z) - \hat{T}_{\hat{\mathfrak{h}}(\s)} (z),  \quad\s = 0, \dots, N_c-1
\\ 
\hat{T}_{\hgs} (z) = \frac{\eta^{ab}}{2k + Q_{\g}} \sum_j \frac{1}{f_j(\s)} \sum_{\hat{j} = 0}^{f_j(\s) -1} : \hj_{\hat{j} a j}(z,\s) \hj_{-\hat{j}, bj}(z,\s) :\, ,\quad a, b = 1, \dots \dim \g\, \\
\hat{T}_{\hhs}(z)  = \frac{\eta^{ab}}{2Kk + Q_\g} : \hj_a^{\text{diag}}(z,\s) \hj_b^{\text{diag}}(z,\s) : \;\;  =  \frac{\eta^{ab}}{2Kk + Q_\g} \sum_{j,l}: \hj_{0aj}(z,\s) \hj_{0bl}(z,\s) :\\
\hat{c}_{\hgs/\hhs}=c_{g/h}
 = K x_{\g}\dim \gfrak(\frac{1}{x_{\g}+\tilde{h}_{\g}}
-\frac{1}{Kx_{\g}+\tilde{h}_{\g}}) \\
\hat{\Delta}_0(\s)=\frac{c_{\g}}{24}(K-\sum_j \frac{1}{f_j(\s)}), \quad c_\g = \frac{x_\g \dim \g}{x_\g + \tilde{h}_\g} \label{cfw}\\
\hat{T}_{\frac{\hgs}{\hhs}}(z) \hj_{a}^{\text{diag}}(w,\s) = {\cal O}((z-w)^0), \quad a = 1, \dots, \dim \g.
\end{gather}
\end{subequations}
In this case, the main characteristics of  $\hat{T}_{\hgs/\hhs}$ are controlled by the properties of the twisted affine-Sugawara construction $\hat{T}_{\hhs}$ on $\hhs$: In the first place, since the currents   $\hj^{\text{diag}}(z, \s)$ of $\hhs$ have trivial monodromy, then $\hat{T}_\hhs(z)$  contributes nothing to the ground state conformal weight in Eq. (\ref{cfw}). In the second place, the diagonal permutation coset orbifolds are {\em interacting} coset orbifolds because the stress tensor $\hat{T}_{\hhs}(z)$ and hence the coset stress tensor   $\hat{T}_{\hgs/\hhs}$ show interaction among the commuting $j$ and $l$ subsystems.


Using the twisted affine embedding $\hj^{\text{diag}} \subset \hj_\hgs$ in (\ref{fs1}) and the affine algebra $\hhs$ in (\ref{secondstar}), the recipe (\ref{recipe}) gives the twisted representation matrices $\st_{h(\s)} = \{\st^{\text{diag}}_a(T,\s)\}$ and their orbifold Lie subalgebra $h(\s)$
\begin{subequations}
\begin{gather}
\st_a^{\text{diag}}(T, \s) \equiv \sum_j \st_{0aj}(T, \s)  \label{Tdg}\\ 
\big[ \st_a^{\text{diag}}(T, \s), \st_b^{\text{diag}}(T, \s) \big] = i f_{ab}{}^c \st_c^{\text{diag}}(T,\s), \quad a,b,c=1,\dots, \dim \g.
\end{gather}
\end{subequations}
Moreover, we may use the factorized form of $\st(T,\s) = T t(\s)$ in (\ref{formofTperm}) to obtain a more explicit form of $\st_a^{\text{diag}}(T,\s)$ 
\be\label{tdig}
\st_a^{\text{diag}}(T, \s) = T_a \otimes \idop(\s)\, , \qquad [T_a, T_b] = if_{ab}{}^c T_c, \quad \idop(\s)_{\hat{j} j}{}^{\hat{l}l} = \delta_j{}^l \delta_{\hat{j} - \hat{l}, 0 \rmod f_j(\s)} 
\e
where $\{T_a\}$ is an irreducible representation of $\gfrak$. In fact the form given for $\s=0$ in (\ref{tdig})
\be
T_a^{\text{diag}}\equiv \st_a^{\text{diag}}(T,0) = \bigoplus_{I=0}^{K-1} T_a^I, \quad T_a^I \cong T_a
\e
is nothing but the untwisted Lie embedding $\{T_h\} \subset \{T_g\}$ implied by the affine embedding (\ref{Jdiag}). More generally, the form of $\st^{\text{diag}}(T,\s)$ in (\ref{tdig}) gives the subgroup orbifold elements $\hat{h}$ for the diagonal permutation coset orbifolds
\begin{subequations}
\begin{eqnarray}
\hat{h}(\st, \xi, t, \s) 
 & =&  e^{i \sum_a \hat{\beta}^{0a}(\xi, t, \s) \st_{a}^{\text{diag}} (T,\s)}\label{1010a}\\
&=& e^{i \sum_{a,j} \hat{\beta}^{0 a}(\xi, t, \s) \st_{0 a j} (T,\s)}\label{h2ex} \label{1010b} \\
& =&  \hat{\hat{h}}(T, \xi, t, \s)\otimes \idop(\s) \; \in \widehat{\HH}(\s) \label{h2exa}
\end{eqnarray}
\vspace{-.3in}
\begin{gather}
\hat{\hat{h}}(T, \xi, t, \s) \equiv  e^{i \sum_a \hat{\beta}^{0a}(\xi, t,\s) T_a}, \quad \forall \; \hat{\beta}^{0a}(\xi, t,\s),  \quad \forall \;T\; \text{for each }\s = 0, \dots, N_c-1 \label{hdiag}.
\end{gather}
\end{subequations}
To obtain the form of $\hat{h}$ in (\ref{1010a}), we used the general form (\ref{antriv}) of the subgroup orbifold element, now with $\st_{h(\s)} = \{\st^{\text{diag}}\}$. The form of $\hat{h}$ in (\ref{1010b}) follows from (\ref{Tdg}), (\ref{1010a}), and comparison of this form with the general form of the group orbifold element in (\ref{g_ex_perm}) tells us that
\be
\hat{\beta}^{\hat{j} aj}(\xi, t, \s) = \delta_{\hat{j}, 0 \rmod f_j(\s)} \hat{\beta}^{0a}(\xi, t, \s) ,  \quad\s = 0, \dots, N_c-1
\e
is the restriction to the infinite-dimensional Lie subalgebra $\hat{h}(\s)$ of the general tangent space coordinates $\hat{\beta}^{\hat{j} a j}(\xi,t,\s)$ on the ambient infinite-dimensional Lie algebra $\hat{g}(\s)$.
The monodromy of $\hat{\beta}^{0a}(\xi,t,\s)$ and hence the monodromies of $\hat{h}(\st, \xi, t,\s)$ and its associated quantities are trivial
\begin{subequations}
\begin{gather}
\hat{\beta}^{0a}(\xi + 2\pi, t, \s) = \hat{\beta}^{0a}(\xi, t,\s), \quad \hat{h}(\st, \xi+2\pi, t,\s) = \hat{h}(\st, \xi, t, \s)\label{trivmono} \\
 \hat{A}_\pm(\st, \xi+2\pi, t,\s) = \hat{A}_\pm(\st, \xi, t, \s), \quad  \hat{\psi}(\st, \xi+2\pi, t,\s) = \hat{\psi}(\st, \xi, t, \s)
\end{gather}
\end{subequations}
in parallel to the trivial monodromy of $\hj^{\text{diag}}(z,\s)$. 

The trivial monodromies in (\ref{trivmono}) are in fact consistent with the apparently non-trivial monodromy of the more general subgroup orbifold element $\hat{h}(\st, \xi, t, \s)$ given in (\ref{hmonop}). To see this, consider the special case 
\begin{gather}
E_0(\s) = 1, \quad
\st_{0aj}(T,\s) = E(T, \s) \st_{0aj}(T, \s) E(T, \s)^*\label{Tpermselec}
\end{gather} 
 of the $\st$-selection rule in  (\ref{sel-rule-for-st-1}). Then the identity
\begin{gather}
 \hat{h}(\st, \xi+2\pi, t,\s) = E(T, \sigma) \hat{h}(\st, \xi, t, \s) E(T,\s)^* = \hat{h}(\st, \xi, t, \s)\label{trmono}
\end{gather}
follows from (\ref{Tpermselec}) and the last form of $\hat{h}(\st, \xi, t,\s)$ in (\ref{h2ex}).
The identity in (\ref{trmono}) is an extreme example of the change in appearance of Eq.\,(\ref{hmonE}) when the support of the subgroup orbifold elements is taken into account. Other examples of this type are found in Sec.\,\ref{sect:other_ex}.

Eq.\,(\ref{h2exa}) shows that these subgroup orbifold elements and hence the twisted gauge fields are also direct sums
\begin{subequations}
\begin{gather}
\hat{h}(\st, \xi, t,\s)_{\hat{j}\alpha j}{}^{\hat{j}'\beta j'}\! = \delta_j{}^{j'} \hat{h}_j(\st, \xi, t,\s)_{\hat{j} \alpha}{}^{\hat{j}' \beta} , \\
\hat{h}_j(\st, \xi, t, \s) = \tau_0(j,\s) \hat{\hat{h}}(T, \xi, t,\s)\\
\hat{A}_\pm(\st, \xi, t,\s)_{\hat{j}\alpha j}{}^{\hat{j}'\beta j'} = \delta_j{}^{j'}
\hat{A}_{\pm,j}(\st, \xi, t,\s)_{\hat{j}\alpha }{}^{\hat{j}'\beta}\\
\hat{A}_{\pm,j}(\st, \xi, t,\s)_{\hat{j}\alpha}{}^{\hat{j}'\beta} = \tau_0(j,\s)_{\hat{j}}{}^{\hat{j}'}  A_\pm(T, \xi, t,\s)_\alpha{}^\beta , \quad A_{\pm}(T, \xi, t, \s)_\alpha{}^\beta \equiv B_\pm{}^a(\xi,t,\s) (T_a)_\alpha{}^\beta\\
\tau_0(j,\s)_{\hat{j}}{}^{\hat{j}'} = \delta_{\hat{j} - \hat{j}', 0 \rmod f_j(\s)}
\end{gather}
\end{subequations}
where the explicit form of $B_\pm$ and hence the reduced matrix gauge fields $(A_\pm)_\alpha{}^\beta$ can be determined from (\ref{blocks_1}) and (\ref{blocks_2}). In this case the matrix $\tau_0(j,\s)$ and hence the blocks $\hat{h}_j$ and $\hat{A}_{\pm , j}$ are essentially independent of $j$, and the permutation coset orbifold action (\ref{moreactions}) takes the form 
\begin{subequations} \label{diag_act}
\begin{align}
 \widehat{S}_{\hg(\s)/\hat{h}(\s)}[\M, \hg, \hat{A}_{\pm}] & =  \sum_j \Big[ \widehat{S}_{\s,j} [ \hat{g}_j] + \widehat{S}_{\s,j}[\hg_j, \hat{A}_{\pm,j}]\Big]\\ 
= \sum_j \bigg\{ \widehat{S}_{\s,j}[\hg_j] &- \frac{k}{4\pi y(T)} \int d^2\xi   \sum_{\hat{j}=0}^{f_j(\s)-1} \sum_{\alpha, \beta=1}^{\dim T} i \Big( (\hat{g}_j^{-1} \partial_+\hat{g}_j)_{\hat{j} \alpha}{}^{\hat{j} \beta} ({A}_-)_{\beta}{}^{\alpha} -  ({A}_+)_{\alpha}{}^{\beta} (\partial_-\hat{g}_j \hat{g}_j^{-1})_{\hat{j} \beta}{}^{\hat{j} \alpha}\Big)\label{ff10} \\
- \sum_{\hat{j}, \hat{j}' = 0}^{f_j(\s) -1}& \sum_{\alpha, \beta, \gamma, \delta=1}^{\dim T} (\hat{g}_j)_{\hat{j} \alpha}{}^{\hat{j}' \beta} ({A}_+)_{\beta}{}^{\gamma} (\hat{g}_j)_{\hat{j}' \gamma}{}^{\hat{j} \delta}({A}_-)_{\delta}{}^{\alpha} \Big)
\bigg\}  - \frac{Kk}{4\pi y(T)} \int d^2\xi \sum_{\alpha, \beta=1}^{\dim T} ({A}_{+})_\alpha{}^{\beta}( {A}_-)_\beta{}^\alpha\nn
\end{align}
\begin{gather}
 \hat{g}_j\equiv \hat{g}_j(\st(T,\s), \xi, t,\s), \quad {A}_\pm  \equiv {A}_\pm (T, \xi, t,\s), \quad \forall\, T \quad \text{for each } \srange
\end{gather}
\end{subequations}
for the diagonal permutation coset orbifolds, where $\widehat{S}_{\s,j} [ \hat{g}_j]$ is given in (\ref{S_j}). To obtain the last term in (\ref{ff10}) we used the identity 
\be
\sum_j \sum_{\hat{j}, \hat{j}' = 0}^{f_j(\s) -1} \tau_0(j,\s)_{\hat{j}}{}^{\hat{j}'} \tau_0(j,\s)_{\hat{j}'}{}^{\hat{j}} =\sum_{j}f_j(\s) =  K
\e
where $K$ is defined in (\ref{Kdef}).  For this action, the reduced matrix gauge fields $A_\pm$ couple to all blocks $\hg_j$, so integration over $A_\pm$ gives interactions among all $\hat{g}_j$ --- consistent with the interacting form of $\hat{T}_{\hgs/\hhs}$ in (\ref{Texa1}). Such interaction among the blocks $\hat{g}_j$ will be found in the action formulation of all interacting coset permutation orbifolds, consistent with the interacting form of their stress tensors.



\section{Other Examples}\label{sect:other_ex}
In this section we survey the literature to find  twisted current-algebraic embeddings which can be used with the recipe of Sec.\,\ref{sect:twistedsubal} to obtain the tangent space description of $\hg$ and $\hat{h}$ in each sector of a number of other coset orbifolds. For brevity, substitution of this data into the general coset orbifold action (\ref{empty_+}) is left as an exercise for the reader.

\subsection{Sub-Example: A Large Class of Cyclic Permutation Coset Orbifolds}
In Ref.\,\cite{Evslin:1999ve}, the twisted current-algebraic formulation is given for a large class of interacting cyclic permutation coset orbifolds
\be
\frac{\Big(\frac{\bigoplus_{I=0}^{\lambda-1}\g^I}{h(\eta, \Phi)} \Big)}{\Z_\l},\quad \g^I\cong \g,  \quad \Z_\l \subset Aut(g), \quad \Z_\l \subset Aut\big(h(\eta, \Phi)\big)
\label{anothex}
\e
where $h=h(\eta, \Phi) \subset g = \bigoplus_{I=0}^{\lambda-1}\g^I$ is a large class of {\em $\Z_\l$-covariant subalgebras} of $g$.\f{The subalgebra $h(\eta, \Phi)$ is specified by any integer $\eta$ which divides $\l$, and a collection $\Phi$ of arbitrary functions $\{ \phi(\alpha_i): 0 \leq \phi(\alpha_i) \leq \frac{\lambda}{\eta}-1, \; i = 1, \dots, \dim \g\}$ on the simple roots $\{\alpha_i\}$ of $\g$.} 

In particular the diagonal cyclic permutation coset orbifold which is included in the discussion of  Sec.\,\ref{sect:diag}
\be
h(1, \{0\}) = \g_{\text{diag}}
,\quad \frac{\left(\frac{\bigoplus_{K=0}^{\lambda-1} \g^I}{\g_{\text{diag}}}\right)}{\Z_\l} \subset \frac{\frac{\bigoplus_{K=0}^{\lambda-1} \g^I}{\g_{\text{diag}}}}{H(\text{permutation})}
\e
is also included in this class as the special case with $\eta = 1$ and $\Phi = \{0\}$. In fact,  $h(1, \{0\})$ is the only case in this class for which the $\Z_\l$-covariant subalgebra $h(\eta, \Phi) \subset g$ is a $\Z_\l$-invariant subalgebra for all $h_\s \in \Z_\l$, $\s = 0, \dots, \lambda-1$. Another special case included in this class is the cyclic coset orbifold
\begin{subequations}
\begin{gather}
h(2,\{0\}) = \g_{2k} \oplus \g_{2k}, \quad \frac{\frac{\g_{k} \oplus \g_{k}\oplus \g_{k} \oplus \g_{k} }{\g_{2k} \oplus \g_{2k}}}{\Z_4}\\
J_a^{R=0}(z) = J_{a0}(z) + J_{a2}(z), \quad J_a^{R=1}(z) = J_{a1}(z) + J_{a3}(z), \quad a=1, \dots, \dim \g \label{blobb}
\end{gather}
\end{subequations}
where the $\Z_4$-covariant affinization of $h(2, \{0\})$ is generated by the currents in  (\ref{blobb}). This orbifold has four sectors but  $h(2, \{0\})$ is a $\Z_4$-invariant subalgebra only in sector $\s=2$ with $\rho(2) =2$.

For all the cyclic coset orbifolds in (\ref{anothex}), the embedding (in the Cartan-Weyl basis of $\g$) of the twisted $h$ currents $\hj_{\hhs}$ and the twisted $h$ current algebra $\hhs$ are given respectively  in Eqs. (3.14) and (3.19) of  Ref.\,\cite{Evslin:1999ve}. According to the recipe of Sec.\,\ref{sect:twistedsubal}, we may use these results and the translation dictionary (see Eq.\,(7.5a))
\begin{subequations}
\begin{gather}
(\text{Ref.\,\cite{Evslin:1999ve}})\quad r,  \hj^{(\hat{r})}_{a(j)}(z) \rightarrow \hat{j}, \hj_{\hat{j} aj}(z) \quad \text{(here)}\\
a=1, \dots, \dim \g, \quad \bar{\hat{j}} = 0, \dots, \r(\s)-1, \quad j = 0, \dots, \frac{\l}{\rho(\s)}-1, \quad \s =0, \dots, \l -1\\
a = (A,\alpha), \quad A = 1, ..., \text{rank}\g, \qquad \alpha \in \Delta(\g)
\end{gather}
\end{subequations}
to read off the embedding of the twisted representation matrices $\st_{h(\s)}$ in $\st_{g(\s)}$ and the orbifold Lie algebra $h(\s)$ of the twisted representation matrices. The results are
\begin{subequations}\label{cwex}
\begin{gather}
\st_A^{R,(r)}=T_A\sum_{j=0}^{\frac{\l}{\r(\s)\mu}-1}e^{\frac{2\pi i  jrN(\s)P}{\eta/\mu}}t_{\frac{\rho(\s)\mu}{\eta}r, \mu j+R}(\s), \quad A = 1, ..., \text{rank}\g \\
\st_\a^{R,(r)}=T_\alpha \sum_{j=0}^{\frac{\l}{\r(\s)\mu}-1}e^{\frac{2 \pi i  j(N(\s)r-\phi(\alpha))P}{\eta/\mu}}e^{\frac{2 \pi i  j\phi(\alpha)}{\l/\mu}}
t_{\rho(\s)(\frac{\mu}{\eta}r+\frac{\s}{\l}\phi(\a)), \mu j+R}(\s), \qquad \forall \, \alpha \in \Delta(\g)
\end{gather}
\begin{gather}
[\st_A^{R, (r)}, \st_B^{S, (s)} ] = 0, \qquad 
[\st_A^{R, (r)}, \st_\alpha^{S, (s)} ] = \delta_{RS} \alpha_A \st_\alpha^{R, (r+s)}\\
[\st_\alpha^{R, (r)}, \st_\beta^{S, (s)} ] = \delta_{RS}
\begin{cases} 
N(\alpha, \beta) \st^{R, (r+s)}_{\alpha + \beta} & \text{if } \alpha + \beta \in \Delta(\g) \\
\alpha \cdot \st^{R, (r+s)} & \text{if } \alpha + \beta = 0 \\
 0 & \text{otherwise}
\end{cases}\\
R, S = 0, \dots, \mu -1, \qquad r,s = 0, \dots, \frac{\eta}{\mu} - 1, \qquad \s = 0, \dots, \lambda-1 
\end{gather}
\end{subequations}
where $T_a = (T_A, T_\alpha)$ are the matrix irreps of $\g$ in the Cartan-Weyl basis. To obtain (\ref{cwex}) from Ref.\,\cite{Evslin:1999ve} we also used the factorized form of the twisted representation matrices
\begin{subequations}
\begin{gather}
\st_{\hat{j} A j}(T,\s)  = T_A t_{\hat{j} j}(\s), \quad \st_{\hat{j} \alpha  j}(T,\s)  = T_\alpha t_{\hat{j} j}(\s), \quad  t_{\hat{j} j}(\s)_{\hat{l} l}{}^{\hat{m} m} = \delta_{jl} \delta_l{}^m \delta_{\hat{j} + \hat{l} - \hat{m}, 0 \rmod \rho(\s)} \\
\quad a = (A, \alpha), \quad \hat{j}, \hat{l}, \hat{m} = 0, \dots, \rho(\s) -1
\end{gather} 
\end{subequations}
which are a special case of the result given for all WZW permutation orbifolds in Eq.\,(\ref{formofTperm}). 
The integers $N(\s), \mu$ and $\phi(\alpha)$ are defined in Ref.\,\cite{Evslin:1999ve}.

This gives the tangent space description and monodromies of the group orbifold elements $\hat{g}$ and the subgroup orbifold elements $\hat{h}$ 
\begin{subequations}
\begin{gather}
\hat{g} (\st, \xi, t,\s)  = e^{i  \sum_{j=0}^{\frac{\lambda}{\rho(\s)}-1} \sum_{\hat{j}=0}^{\rho(\s)-1}\sum_{a=1}^{\dim \g} \hat{\beta}^{\hat{j}aj}(\xi, t, \s)\st_{\hat{j}aj}(\s)}, \quad \forall \;\hat{\beta}^{\hat{j}aj}(\xi, t, \s),  \quad\s = 0, \dots, N_c-1 \\
\hat{h} (\st, \xi, t,\s)  = e^{i \sum_{r=0}^{\frac{\eta}{\mu} -1} \sum_{R=0}^{\mu-1} \big(  \sum_{\alpha \in \Delta(\g)} \hat{\beta}^{r \alpha R}(\xi, t, \s) \st_\alpha^{R,(r)} + \sum_{A=1}^{\text{rank} \g} \hat{\beta}^{r AR}(\xi, t,\s) \st_A^{R, (r)} \big) } , \quad \forall \;\hat{\beta}^{raR}(\xi, t, \s)\\
 \hat{O}(\st, \xi+2\pi, t,\s)_{\hat{j} \gamma j}{}^{\hat{l} \delta l} = e^{-2 \pi i \frac{ \hat{j} -\hat{l}}{\rho(\s)}} \hat{O}(\st, \xi, t, \s)_{\hat{j} \gamma j}{}^{\hat{l} \delta l}, \quad \hat{O} = \hat{g} \text{ or } \hat{h}, \quad \gamma, \delta = 1, \dots, \dim T
\end{gather}
\end{subequations}
in all sectors of each cyclic coset orbifold in this class.

\subsection{Example: A Class of Inner-Automorphic Coset Orbifolds}
The general inner-automorphic WZW orbifold on simple $g$
\be
\frac{A_g(H(d))}{H(d)}, \quad H(d) \subset \text{Lie}G \subset Aut(g)
\e
was considered at the operator level in Refs.\,\cite{Halpern:2000vj,deBoer:MBH01}, and Ref.\,\cite{deBoer:MBH01} also considered these orbifolds at the action level. In this case the action of $H(d)$ in the symmetric CFT $A_g(H(d))$ is given in the Cartan-Weyl basis as
\begin{subequations}
\begin{gather}
H_A(\xi, t)'  = H_A(\xi, t), \quad E_\alpha(\xi, t)' = e^{2\pi i \s \alpha \cdot d} E_\alpha(\xi, t) \\
\quad A = 1, \dots, \text{rank} g, \quad \alpha \in \Delta(g), \quad \srange,  \quad N_c = \rho(1) 
\end{gather}
\end{subequations} 
where Cartan $g$ $\subset g$ is an $H(d)$-covariant subalgebra of $g$, which is an invariant subalgebra in this case. The vector $d$ is chosen  \cite{FH} so that $H(d)$ is a group of finite order. Moreover it is known \cite{Halpern:2000vj,deBoer:MBH01} for the inner-automorphic WZW orbifolds that 
\begin{subequations}
\begin{gather}
n_A = 0, \quad \frac{n_\alpha(r)}{\r(\s)} = -\s \alpha \cdot d, \quad \frac{N(r)}{R(\s)} = \s \l \cdot d,  \quad\s = 0, \dots, \rho(1)-1 \\
\st(T, \s) = T, \quad {\cal M}(\st, \s) = M(k,T) = \frac{k}{y(T)} \idop\\
 (T_A)_\l{}^{\l'} \equiv \;\langle T,\l | H_A(0) | T,\l' \rangle = \l_A\d_{\l,\l'} ,\quad
    (T_\a)_\l{}^{\l'} \equiv \;\langle T,\l | E_\a(0) | T,\l' \rangle
    \propto\; \d_{\l,\l'+\a} \label{rep-matrices-of-inner-auto}\\
e^{-2\pi i \s (\l-\l')\cdot d} (T_A)_{\l}{}^{\l'} = (T_A)_{\l}{}^{\l'} ,
 \quad  e^{-2\pi i \s (\l-\l')\cdot d} (T_\a)_{\l}{}^{\l'} =
 e^{-2\pi i \s \a\cdot d} (T_\a)_{\l}{}^{\l'} \label{eT=T}\\
A= 1, \dots, \text{rank}g, \quad \alpha \in \Delta(g)
\end{gather}
\end{subequations}
where $n(r)$ and $N(r)$ are the spectral integers of the $H$-eigenvalue problem and the extended $H$-eigenvalue problem respectively, $\{\l_A\}$ are the weights of irrep $T$ and $T_a= (T_A, T_\alpha)$ are the representation matrices of $g$ in the weight basis. The explicit form of the inner-automorphically twisted affine Lie algebra, the twisted affine Sugawara construction and the twisted affine primary fields are given for sector $\s$ of $A_g(H(d))/H(d)$ in Refs.\,[3,11]. In particular, the monodromies of the twisted currents of sector $\s$
\be\label{lsts}
\hat{H}_A(\xi+2\pi, t) = \hat{H}_A(\xi, t), \quad \hat{E}_\alpha(\xi+2\pi, t) = e^{2 \pi i \s \alpha \cdot d} \hat{E}_\alpha(\xi, t)
\e
follow from the data in Eq.\,(11.10a).

Then we may read from Ref.\,\cite{deBoer:MBH01} what we need to know about the large class of inner-automorphic coset orbifolds
\be
\frac{A_{\frac{g}{\text{Cartan}\, g}}(H(d))}{H(d)}, \quad H(d) \subset Aut(g), \quad H(d) \subset Aut(\text{Cartan}\, g).
\e 
In particular, we find in Ref.\,\cite{deBoer:MBH01} the form and monodromies of the group orbifold elements $\hat{g}$ 
\begin{subequations}
\begin{gather}
\hat{g}(T, \xi, t,\s) = e^{i( \sum_{A=1}^{\text{rank}g}\hat{\beta}^A(\xi, t, \s) T_A + \sum_{\alpha \in \Delta(g)}\hat{\beta}^\alpha(\xi, t,\s) T_\alpha )}, \quad \forall \; \hat{\beta}^a(\xi, t, \s),  \quad\s = 0, \dots, \rho(1)-1 \\
\hat{\beta}^A(T, \xi+2\pi, t, \s) = \hat{\beta}^A(T, \xi, t, \s), \quad \hat{\beta}^\alpha(T, \xi+2\pi, t, \s) =  \hat{\beta}^\alpha(T, \xi, t, \s) e^{-2 \pi i \s \alpha \cdot d}\label{monob}\\
\hat{g}(T, \xi+2\pi, t, \s)_\lambda{}^{\lambda'} =  e^{-2\pi i \s d \cdot (\l - \l')} \hat{g}(T, \xi, t, \s)_\lambda{}^{\lambda'}\label{monohib} 
\end{gather}
\end{subequations}
of the WZW orbifold $A_g(H(d))/H(d)$. The $\st$-selection rules in (\ref{eT=T}) can be used to verify the consistency of the monodromy of the twisted tangent space coordinates $\{\hat{\beta}\}$  with the monodromy of the full group orbifold elements. One sees here the orbifold Lie subalgebra $h(\s)$ 
\be
[T_A, T_B ] = 0, \quad A,B = 1, \dots, \text{rank} g
\e
generated by the twisted representation matrices $\st_{h(\s)} = \{T_A\}$. The orbifold Lie subalgebra $h(\s)$ is the action analogue of the affine Cartan subalgebra $\hhs$ generated by the modes of $\hj_{\hhs} = \{\hj_A\}$ in (\ref{lsts}). Then we know that the corresponding subgroup orbifold elements satisfy
\begin{subequations}
\begin{gather}
\hat{h}(T, \xi, t, \s) = e^{i\sum_{A=1}^{\text{rank} g} \hat{\beta}^A(\xi, t, \s) T_A}\\
\hat{h}(T, \xi+2\pi, t, \s)_\lambda{}^{\lambda'} = e^{-2\pi i \s d \cdot (\l - \l')}  \hat{h}(T, \xi, t, \s)_\lambda{}^{\lambda'} = \hat{h}(T, \xi, t, \s)_\lambda{}^{\lambda'}\label{monohia}
\end{gather}
\end{subequations}
where we have used the first selection rule in (11.10d) to verify the consistency of the two parts of (\ref{monohia}).
The subgroup orbifold elements have trivial monodromy in parallel with the trivial monodromy of the affine Cartan subalgebra $\hhs$.



\subsection{Example: A Class of Charge Conjugation Coset Orbifolds} 
The outer-automorphic charge conjugation orbifold on invariant level $x$ of $\mathfrak{su}(n)$ 
\be
\frac{A_{\mathfrak{su}(n)}(\Z_2)}{\Z_2} \equiv \frac{\mathfrak{su}(n)_x}{\Z_2}, \quad n\ge 3
\e
was studied at the operator and the action level in Ref.\,\cite{MBHObers}. Here the non-trivial element of  $H=\Z_2$ is charge conjugation, which exchanges each representation of $\mathfrak{su}(n)$ with its charge conjugate representation, and there is an irregularly-embedded $\Z_2$-covariant $\mathfrak{so}(n)$ subalgebra
\be
\mathfrak{so}(n)_{2\tau x} \subset \mathfrak{su}(n)_x , \quad \tau = \begin{cases} 2  & \text{for } n = 3 \\ 1 & \text{for } n \ge 4 \end{cases} 
\e
which in this case is an invariant subalgebra under the action of charge conjugation. 

Then one may read off what we need to know about the single twisted sector $\s =1$ of each charge conjugation coset orbifold 
\be
\frac{\frac{\mathfrak{su}(n)_x}{\mathfrak{so}(n)_{2\tau x}}}{\Z_2}\;,\quad  n \ge 3
\e
from the data given in Sec.\,3.6 of  Ref.\,\cite{MBHObers}. In particular, we find the tangent space description  and monodromies of the group orbifold elements $\hat{g}$
\begin{subequations}\label{lastex}
\begin{gather}
\hg(\st(T), \xi, t) = \exp \left(i\left(\begin{array}{cc}
\hat{\beta}^{0A}(\xi,t) T_A^{(c)} & \hat{\beta}^{1I}(\xi,t) T_I^{(c)}\\
\hat{\beta}^{1I}(\xi,t) T_I^{(c)} & \hat{\beta}^{0A}(\xi,t) T_A^{(c)}
\end{array} \right) \right), \quad 
A \in \mathfrak{so}(n), \quad I \in \frac{\mathfrak{su}(n)}{\mathfrak{so}(n)}\\
\hat{\beta}^{0A}(\xi+ 2\pi, t) = \hat{\beta}^{0A}(\xi, t), \quad 
\hat{\beta}^{1I}(\xi+ 2\pi, t) = -\hat{\beta}^{1I}(\xi, t) \\
\hat{g}(\st(T), \xi+2\pi, t) = E(T) \hat{g}(\st, \xi, t) E(T)^*, \quad E(T)=  i\left(\begin{array}{cc} 1 & 0 \\ 0 & -1 \end{array} \right)
\end{gather}
\end{subequations}
in the twisted sector of each charge conjugation orbifold $\mathfrak{su}(n)_x/\Z_2$. Here one sees the orbifold Lie subalgebra $h = \mathfrak{so}(n)$
\be
[T_A, T_B] = i f_{AB}{}^CT_C, \quad A,B,C \in \mathfrak{so}(n)
\e
generated by the matrices $\st_h = \{T_A\}$. Therefore the subgroup orbifold elements have the form
\begin{subequations}\label{lte}
\begin{gather}
\hat{h}(\st(T), \xi, t) = \exp \left(i \left(\begin{array}{cc}
\hat{\beta}^{0A}(\xi,t) T_A^{(c)} &  0 \\
0 & \hat{\beta}^{0A}(\xi,t) T_A^{(c)}
\end{array} \right)\right) \\
\hat{h}(\st(T), \xi+2\pi, t) = E(T) \hat{h}(\st(T), \xi, t) E(T)^* = \hat{h}(\st(T), \xi, t)
\end{gather}
\end{subequations}
in the twisted sector of these coset orbifolds. 
The results (\ref{lastex}) and (\ref{lte}) apply when the untwisted representation  $T = T^{(c)}$ is any complex irrep of $\mathfrak{su}(n)$, such as the fundamental representation, and the corresponding results for real irreps of $\mathfrak{su}(n)$ are also easily read from Ref.\,\cite{MBHObers}. 

More generally, we know that $H$-covariant subalgebras $h\subset g$ map by local isomorphisms (see Eq.\,(\ref{lim})) onto twisted subalgebras $\hhs \subset \hgs$. Then  the data for the operator and the action formulation of any given coset orbifold $A_{g/h}(H)/H$ can always be found, as we have done throughout this section, in the details of the twisted current algebra $\hgs$ of the WZW orbifold $A_g(H)/H$.

\vspace{0.5in}
\noindent
{\bf Acknowledgement} 

\noindent
For helpful discussions, we thank J. de Boer, S. Giusto, C. Helfgott and N. Obers.

M.B.H. is supported in part by the Director, Office of Science, Office of High Energy and Nuclear Physics of the U.S. Department of Energy under Contract DE-AC03-76SF00098 and in part by the National Science Foundation under grant PHY-0098840. F.W. acknowledges financial support by the DAAD, granted by the Gemeinsamen Hochschul\-sonder\-programm III von Bund und L\"andern.


\end{document}